\documentclass[fleqn,usenatbib]{mnras}

\usepackage[T1]{fontenc}

\DeclareRobustCommand{\VAN}[3]{#2}
\let\VANthebibliography\thebibliography
\def\thebibliography{\DeclareRobustCommand{\VAN}[3]{##3}\VANthebibliography}

\usepackage{newtxtext,newtxmath}
\usepackage{ulem}
\hypersetup{linkcolor=red,citecolor=blue,filecolor=cyan,urlcolor=magenta}
\usepackage{amsmath}
\usepackage{tensor}     
\usepackage{graphicx}   
\usepackage{bm}         
\usepackage{xcolor}     
\usepackage{color}      
\usepackage[section]{placeins} 
\usepackage[utf8]{inputenc} 
\usepackage[flushleft]{threeparttable}
\usepackage{float}
\usepackage{orcidlink}
\usepackage{multirow}
\newcommand{\nc}{\newcommand*} 
\nc{\figurewidth}{3.2in}
\nc{\xbar}{\bar{x}}
\nc{\rhoeq}{\rho_{\mathrm{eq}}}
\nc{\zeq}{z_{\mathrm{eq}}}
\nc{\tla}{\tilde{\lambda}}
\nc{\dt}{\delta}
\nc{\Dt}{\Delta}
\nc{\vj}{\vec{j}}
\nc{\vl}{\vec{l}}
\nc{\hx}{\hat{x}}
\nc{\hy}{\hat{y}}
\nc{\bj}{\bm{j}}
\nc{\mJ}{\mathcal{J}}
\nc{\mP}{\mathcal{P}}
\nc{\Msun}{M_\odot}
\nc{\app}{\approx}
\nc{\av}[1]{\langle #1 \rangle}
\nc{\eq}[1]{Eq.~\eqref{#1}}
\nc{\al}{\alpha}
\nc{\Xstar}{X_{\ast}}
\nc{\seq}{\sigma_{\mathrm{eq}}}
\nc{\fpbh}{f_{\mathrm{pbh}}}
\nc{\vth}{\vec{\theta}}
\nc{\vla}{\vec{\lambda}}
\nc{\vd}{\vec{d}}
\nc{\Mmin}{M_{\mathrm{min}}}
\nc{\rmd}{\mathrm{d}}
\nc{\mmin}{{m_{\mathrm{min}}}}
\nc{\mmax}{{m_{\mathrm{max}}}}
\nc{\mR}{\mathcal{R}}
\nc{\tmR}{\tilde{\mathcal{R}}}
\nc{\s}{\sigma}
\nc{\ogw}{\Omega_{\mathrm{GW}}}
\nc{\addref}{[\textcolor{red}{add ref}] }
\nc{\Om}{\Omega}
\nc{\gpcyr}{\mathrm{Gpc}^{-3}\,\mathrm{yr}^{-1}}
\nc{\Eq}[1]{Eq.~\eqref{#1}}
\nc{\Fig}[1]{Fig.~\ref{#1}}
\nc{\Table}[1]{Table~\ref{#1}}
\nc{\lvc}{LIGO/Virgo} 
\nc{\Sec}[1]{Sec.~\ref{#1}}
\nc{\eg}{\textit{e.g.~}}
\nc{\SNR}{\mathrm{SNR}}

\def\({\left(}
\def\){\right)}
\def\[{\left[}
\def\]{\right]}

\def\e{\begin{equation}}
\def\q{\end{equation}}
\def\m{\begin{eqnarray}}
\def\n{\end{eqnarray}}

\newcommand{\hs}{h_{\rm s}}
\newcommand{\hc}{h_{\rm c}}
\newcommand{\ayr}{A_{\rm yr}}
\newcommand{\fyr}{f_{\rm yr}}

\newcommand{\mbh}{M_{\bullet}}

\newcommand{\dist}{D_{\rm c}}

\newcommand{\fgearth}{f}
\newcommand{\fgrest}{f_{\rm r}}

\newcommand{\be}{\begin{equation}}
\newcommand{\ee}{\end{equation}}

\title[Probing Merger Rates of SMBHs \& Galaxies with GWs]{Probing the Merger Rates of Supermassive Black Holes and Galaxies with Gravitational Waves}

\author[Y. Fang $\&$ R. Cai]{
Yun Fang \!\orcidlink{0000-0003-0065-8622}, $^{1,2}$\thanks{E-mail: fangyun@nbu.edu.cn}
Rong-Gen Cai,$^{1,2}$\thanks{E-mail: caironggen@nbu.edu.cn}
\\
$^{1}$Institute of Fundamental Physics and Quantum Technology, Ningbo University, Ningbo, 315211, China\\
$^{2}$School of Physical Science and Technology, Ningbo University, Ningbo, 315211, China
}

\date{Accepted XXX. Received YYY; in original form ZZZ}
\pubyear{2025}
\begin{document}
\label{firstpage}
\pagerange{\pageref{firstpage}--\pageref{lastpage}}
\maketitle

\begin{abstract}

The mergers of galaxies and supermassive black holes (SMBHs) are key drivers of galaxy evolution, contributing to the growth of both galaxies and their central black holes. 
Current and upcoming gravitational wave (GW) detectors—Pulsar Timing Arrays (PTAs), LISA, Taiji, and Tianqin—offer unique access to these processes by observing GW signals from SMBH binaries. We present a framework to infer galaxy and SMBH merger rates by combining mock LISA detections of SMBH mergers with PTA constraints on the stochastic GW background, while incorporating observational uncertainties in stellar mass functions and $M_\bullet$–$M_*$ relations. We find that the number of LISA-detected events and their joint distribution in mass and redshift are key to constraining merger rates—datasets with around forty events yield results consistent with galaxy pair observations, whereas limited event counts lead to biases at high redshift. Including PTA data further reduces parameter uncertainties. 
Our method also effectively constrains the delay time between galaxy and SMBH mergers, with longer delays suppressing high-redshift SMBH merger rates and shifting mass growth from mergers to accretion. According to our mock analysis, the models with delay times longer than $0.5\text{Gyr}$ ($0.8\text{Gyr}$), accretion becomes the primary driver of SMBH mass growth beyond $z \sim 6$ ($4$). In contrast, the SMBH occupation fraction at $z>3$ remains poorly constrained due to its degeneracies with delay time and the galaxy merger rate. These findings highlight both the promise and limitations of using GW observations to probe the coevolution of galaxies and SMBHs. 
            
\end{abstract}


\section{Introduction}
Supermassive black holes (SMBHs) are ubiquitous in the centers of massive galaxies. Observations reveal that SMBHs are intimately linked to the properties of their host galaxies, as demonstrated by scaling relationships like the $M_{\bullet}-\sigma$ and $M_{\bullet}-M_{*}$ relations \citep[][]{Ferrarese_2000, Gebhardt:2000fk, Kormendy:2013dxa}, which link the mass of the SMBH $M_{\bullet}$  to the velocity dispersion $\sigma$ of the galaxy bulge and the galaxy stellar mass ($M_{*}$). 
This intimate relationship suggests an co-evolution of SMBHs and their host galaxies throughout the universe.

Galaxies are hierarchically assembled through the mergers of dark matter halos. When two galaxies merge, their central SMBHs form pairs at large separations. Over time, these black holes lose energy via dynamical friction, interactions with gas and stars, and eventually settle into a regime where gravitational radiation dominates \citep[][]{Begelman1980, Yu2002, Escala_2005}. Ultimately, they coalesce into a single, more massive black hole. 
Throughout galaxy evolution, SMBHs assemble their mass through both gas accretion and hierarchical mergers, with accretion generally considered as the dominant process and merger becomes significant for massive SMBHs at low redshifts
 \citep[e.g.][]{Soltan1982, Marconi:2003tg, Hopkins_2006, Volonteri2012Sci, Pacucci_2020}. Understanding the evolutionary history of SMBHs and their coevolution with host galaxies is vital for unraveling the processes driving galaxy formation and evolution, as well as the emergence of the large-scale structure of the universe.

One of the recent spotlight observations comes from the pulsar timing arrays (PTAs): the North American Nanohertz Observatory for Gravitational Waves (NANOGrav) \citep[][]{NANOGrav_2023}, the European PTA (EPTA)  in conjunction with the Indian PTA (InPTA) \citep[][]{EPTA:2023sfo, EPTA:2023fyk}, the Parkes PTA (PPTA) \citep[][]{Reardon_2023}, and the Chinese PTA (CPTA) \citep[][]{CPTA_2023} announced the 
evidence for a signal consistent with a stochastic gravitational wave background (SGWB). 
This SGWB signal is most likely generated by SMBH binaries across the universe during their inspiraling phase at sub-parsec separations. The strain amplitude of the SGWB constrained by PTAs, assuming quasi-circular orbits for SMBH binaries, suggests that SMBH binaries experience no delay or short delay timescale between the merger of SMBHs and galaxies \citep{EPTA:2023xxk}. 
This finding implies that the "final parsec problem" \citep{Milosavljevi__2001,Yu2002, Milosavljevic2003b} may not be a significant barrier. 
A short delay timescale could be explained if the SMBH binaries reside in a triaxial stellar distribution \citep[e.g.][]{Yu2002, Merritt_2004, Holley-Bockelmann:2006gbs, Gualandris2016}, a gas-rich environment \citep[e.g.][]{Armitage2002, Dotti_2006_nucleardisc, Haiman_2009}, or if the SMBHs have undergone multiple mergers \citep{Hoffman_Loeb_2007}.  

Another major highlight comes from recent discoveries by the James Webb Space Telescope (JWST) \citep[][]{ubler2023, Larson2023, Harikane2023, Bogdan:2023ilu, Ding2023Natur, Maiolino:2023bpi, Yue2024, Kocevski2023, Stone2024} and the observations of scaling relationships at redshifts $z > 4$ from a population of SMBHs  \citep[][]{Pacucci_2023, JunyaoLi2024, matthee2024environmental}. Notably, JWST observations suggest that the masses of SMBHs at $4<z<7$ are one to two orders of magnitude larger than those predicted by the local $M_{\bullet}-M_{*}$ relation \citep[][]{Pacucci_2023}, albeit with a larger scatter. The theoretical model has been proposed to explain these JWST findings \citep[][]{Pacucci_2024}. However, there is also speculation that this apparent offset toward higher SMBH masses may be due to observational biases \citep[][]{JunyaoLi2024}.   

Future projects such as the Laser Interferometer Space Antenna (LISA) \citep[][]{eLISA:2013xep, LISA_2017}, Taiji \citep[][]{YueliangWu_taiji, RuanWenHong2018}, and Tianqin \citep[][]{luo2016tianqin} are designed to detect millihertz gravitational waves (GWs) generated by SMBH binary mergers. While both PTAs and LISA-like detectors are able to detect GWs from SMBH binaries, they are complementary in their capabilities. 
PTAs are designed to measure GWs from individual SMBH binaries and/or the SGWB produced by unresolved SMBH binaries, primarily within the mass range of $10^{8}-10^{9}\Msun$. 
In contrast, LISA-like detectors are optimized to detect GWs from merging SMBH binaries with masses in the range of $10^{3}-10^{8}\Msun$.
Given that gravitational radiation from SMBH binary mergers represents some of the most energetic events in the universe, the signals are expected to be so strong that LISA will be able to detect them with high signal to noise ratio, potentially enabling the detection of such sources out to redshifts as high as $z \sim 20-30$ \citep[][]{LISA_2017}. By comparison, for PTAs, the SGWB is dominated by SMBH binaries within $z\sim 3$ \citep[e.g.][]{Sesana2013}, and the detection of individual binaries is typically limited to redshifts of $z\sim1$ \citep[e.g.][]{Sesana2009}.   

Detecting GWs from SMBH binaries is crucial for revealing the formation and evolutionary histories of SMBHs. Such observations enable the estimation of their merger rates \citep[e.g.][]{Klein_2016, Katz_LISA_2020, EPTA:2023xxk, Bi:2023tib}, the differentiation between seeding models \citep[][]{Klein_2016}, the investigation of their evolutionary timescales \citep[e.g.][]{Fang:2022cso, Chen_2023, EPTA:2023xxk}, and determining the role mergers play in SMBH mass assemble \citep[e.g.][]{Pacucci_2020, Valiante2021} as well as in shaping scaling relationships \citep[e.g.][]{Volonteri2009, Shankar2016, Simon_2016}. 
  
  In this work, we estimate the merger rate of SMBHs and galaxies from GW detection with mock LISA GW data and the current PTA detection of SGWB strain \citep[e.g.][]{NANOGrav_2023}. 
 Since direct observations of SMBH pairs, binaries, and their mergers are challenging, SMBH merger rates are typically inferred indirectly from galaxy merger rates, e.g., using the scaling relationships between SMBHs and their host galaxies.
We adopt the local $M_{\bullet}-M_{*}$ scaling relationship from \cite{Kormendy:2013dxa} for $z<4$ and incorporate JWST results \citep[][]{Pacucci_2024} for $z>4$. 
 The galaxy merger rate is determined as the product of the galaxy merger rate per galaxy and the galaxy stellar mass function (GSMF).
The GSMF is well-constrained at low and intermediate redshifts, and we utilize its observational results extending up to $z\sim 10$ \citep[][]{Baldry2012, HuertasCompany2016, Santini2012, McLeod2021, Song2016, Stefanon2021}. 
The galaxy merger rate per galaxy however remains uncertain. 
It is either inferred from electromagnetic observations of galaxy pairs—calculated as the galaxy pair fraction divided by an assumed delay time between pair formation and merger \citep[e.g.][]{Sesana:2012ak, Duncan2019, DuanQiao2024}—or predicted through theoretical models \citep[e.g.][]{Vicente2015, OLeary_2021}. 
We  parameterize the galaxy merger rate per galaxy with hyper-parameters, which, together with the delay time of SMBH mergers and the SMBH occupation fraction, constitute the model parameters. In the process of estimation of model parameters, mock LISA GW data for representative examples are generated, while accounting for model uncertainties from observational uncertainties in both the GSMFs and the $M_{\bullet}-M_{*}$ relationship at different redshift bins. Our discussion focuses on following aspects: 1. The capability of the method developed here to constrain the galaxy merger rate per galaxy, the delay time of SMBH mergers, and the SMBH occupation fraction for low-mass galaxies; 2. A comparison between the galaxy merger rates inferred from GW detections and those derived from galaxy pair observations and cosmological simulations; 3. An analysis of SMBH mass assembly, comparing the contributions from mergers and accretion, as well as the influence of the delay timescale on the SMBH merger rate and mass assembly.  

This paper is organized as follows. In Section \ref{sec_2}, we outline the framework for constructing the merger rate of SMBH binaries. 
Section \ref{sec_3} presents the derivation of the SGWB strain from a population of SMBH binaries. 
In Section \ref{sec_4}, we introduce the population analysis framework for LISA GW data. 
Section \ref{sec_5} is devoted to inferring galaxy and SMBH merger rates from mock LISA data, in combination with PTA constraints. Specifically, we estimate model parameters under two merger rate models in subsections \ref{sec:theoretical_merger_rate}–\ref{infer_merger_rate}, compare galaxy merger rates inferred from GW detection with those from observations of galaxy pairs and simulations in subsection \ref{sec:compare_merger_rate}, estimate the delay time of SMBH binary coalescence in subsection \ref{sec_delay_time}, discuss SMBH mass assembly contributed by mergers and accretion in subsection \ref{sec:SMBH_mass_assembly}, and analyze the SMBH occupation fraction in subsection \ref{sec_occupation}. Finally, we conclude our work in Section \ref{conclusion}.

In the context, we adopt a $\Lambda \text{CDM}$ cosmology with $H_0=70\ \text{km}\ \text{s}^{-1} \text{Mpc}^{-1}$, $\Omega_{\text{M}} = 0.3$, and $\Omega_{\Lambda}=0.7$ at $z=0$. 
  
\section{The merger rate of SMBH binaries}
\label{sec_2}
Galaxy mergers bring together their SMBHs, forming SMBH pairs. During major mergers, dynamical friction works as a main process that drives the SMBHs toward the center of the common nucleus of the newly formed galaxy, where they eventually form a Keplerian binary at separations of a few parsecs  \citep{Begelman1980, Yu2002, Mayer:2007vk, Callegari:2008py, Chapon2013, Pfister2017, Sayeb2021}. The subsequent evolution of the binary orbit depends on interactions with the surrounding environment. Over a characteristic dynamical timescale, the SMBH binary sinks to sub-parsec separations, where gravitational radiation becomes the dominant mechanism driving orbital decay. At this stage, the binary emits GWs in the nanohertz frequency band, contributing to the SGWB detectable by PTAs. Finally, the SMBH binary coalesces due to energy dissipation via gravitational radiation, producing GW signals that could be detected by detectors like LISA.

The merger rate of SMBH binaries, $\frac{d^{3} n}{dz ~dM_{\bullet} ~dq_{\bullet}}$ , is determined by the SMBH mass function, $\Phi_{\bullet} (z, M_{\bullet})$, and the merger rate per SMBH, $\mathcal{R}_{\bullet} (z, M_{\bullet}, q_{\bullet})$,
\be
	\frac{d^{3}n_{\bullet}}{dz ~dM_{\bullet} ~dq_{\bullet}} = \Phi_{\bullet} (z, M_{\bullet}) ~\mathcal{R}_{\bullet} (z, M_{\bullet}, q_{\bullet}) \,, 
	\label{eqn:rate_SMBH_BH}
\ee
where $q_{\bullet}$ is the mass ratio of the merging SMBHs. 

A scaling relation is commonly employed to connect the merger rate of SMBH binaries to the merger rate of galaxies, to be specific,
\begin{eqnarray}
\label{relate_BH_merge_galaxy_merger}
	    \frac{d^{3}n_{\bullet}}{dz ~dM_{\bullet} ~dq_{\bullet}} &=& \left| \Phi_{\rm Gal} ~\mathcal{R}_{\rm Gal} \right|_{\mathrm{Gal} \rightarrow \bullet} \nonumber\\
&=& f_{\text{occ}} \frac{d^{3}n_{\mathrm{Gal} }}{dz ~d\mathcal{X}_{\mathrm{Gal} } ~dq_{\mathrm{Gal} }} \frac{d\mathcal{X}_{\mathrm{Gal}}} {dM_{\bullet} } \frac{dq_{\mathrm{Gal}}}{dq_{\bullet} } \,,
\end{eqnarray}
where $\mathcal{X}_{\mathrm{Gal}} $ denotes the property of host galaxy, such as the stellar mass $M_*$, the budge mass $M_{\rm budge}$, or the velocity dispersion $\sigma$, $q_{\mathrm{Gal}}$ is the mass ratio of the host galaxies , and $f_{\text{occ}}$ is the SMBH occupation fraction of galaxy. 

It has been suggested \citep[][]{Pacucci_2023, Maiolino:2023bpi} that the $M_{\bullet}-\sigma$ relation is more fundamental, as it exhibits smaller scatter and remains largely invariant across redshifts compared to the $M_{\bullet}-M_{*}$ relation, which is subject to greater uncertainty, and hints of evolution with redshift. 
However, the galaxy velocity dispersion function, $\Phi_{\rm Gal}(\sigma)$, is poorly constrained, particularly at high redshifts \citep[][]{Taylor_2022, Matt2023}. 
 In contrast, the GSMF has been properly measured up to $z\sim 10$ \citep[][]{Baldry2012, HuertasCompany2016, Santini2012, McLeod2021, Song2016, Stefanon2021}. 
Since in this work, we consider the detection with LISA which is supposed to measure GW events at high redshift, the $M_{\bullet}-M_{*}$ relation is adopted here.
Consequently, $\mathcal{X}_{\mathrm{Gal}}$ in equation~(\ref{relate_BH_merge_galaxy_merger}) represents $M_{*}$. 

The $M_{\bullet}-M_{*}$ relationship could be parameterized as,
\be \label{scaling_relation}
\log_{10} \left({M_{\bullet}\over \Msun}\right) = \mathcal{N}\left( a + b \log_{10} \left( {M_{*}\over \Msun} \right) ,\epsilon \right) \,,
\ee
where $\mathcal{N}(\mu, \epsilon)$ represents a normal distribution with a mean value of $\mu$  and a scatter of $\epsilon$. We now denote the parameters $\{a, b, \epsilon\}$ as $\Lambda_{M_{\bullet}-M_{*}}$ in the following context. 

 The scatter in the $M_{\bullet}-M_{*}$ relationship has a significant impact on the derived mass function and the merger rate of SMBHs \citep[e.g.][]{Simon_2016, Simon_2023}.  
 Additionally, there is typically a delay time $\tau$ between the merger of the SMBH binary and the merger of their host galaxies. 
  To account for the delay time and the uncertainties arising from the scaling relationship, the merger rate of SMBH binaries is expressed as \citep[e.g.,][]{Fang:2022cso}, 

  \m \label{BH_merger_rate}
\!\! \!\!\! \!\!\! \!\!\!\! \!\!\! \!\!\! \!\! &&\frac{d^{2}n_{\bullet}}{dz ~dM_{\bullet}} (z, M_{*} |{\bf \Lambda}) =\nonumber \\
\!\! \!\!\! \!\!\! \!\!\!\! \!\!\! \!\!\! \!\!&& {dt_{\rm L}\over dz}   
 \!\!\int  \!\!\! 
 \frac{d^{2}n_{\rm Gal}}{dz ~dM_{*}}(t_{\text {L}}\! \!+\!\tau, M_{*} |{\bf \Lambda}_{\text{m}}) f_{\text{occ}}  P_{{\rm delay}} (\tau|{\bf \Lambda}_{\text{d}}) P (M_{\bullet}| M_{*}) d M_{*} d\tau  \,,
  \n
 where $t_{\text {L}}$ is the lookback time, $P (M_{\bullet}| M_{*})$ is the conditional probability determined by equation~(\ref{scaling_relation}), $P_{\text{delay}} (\tau|{\bf \Lambda}_{\text{d}}) $ is an arbitrary distribution of delay time parameterized with ${\bf \Lambda}_{\text{d}}$, and the galaxy merger rate $\frac{d^{2}n_{\rm Gal}}{dz ~dM_{*}}(t_{\text {L}}\! \!+\!\tau,\! M_{*} |{\bf \Lambda}_{\text{dN/dt}})$ is parameterized with ${\bf \Lambda}_{\text{dN/dt}}$. The model of SMBH merger rate is then given by model parameters ${\bf \Lambda}=\{{\bf \Lambda}_{\text{m}}, {\bf \Lambda}_{\text{d}}, f_{\text{occ}}\}$. 
  We neglect the time SMBH binaries spend emitting in the nanohertz band prior to merger. This approximation is due to the fact that this time is typically much smaller than the Hubble time. 

In this work, we focus on major galaxy mergers, restricting the mass ratio $q_{\mathrm{Gal}}$ to the range $[1/4, 1]$. 
The galaxy merger rate is averaged over mass ratios within this range. Consequently, the galaxy merger rate is expressed as:
\begin{align} \label{galaxy_rate2}
\frac{d^{2}n_{\mathrm{Gal} }}{dz ~dM_*} &= {\Phi_{\rm GSMF}(M_*, z)} \mathcal{R}_{\rm Gal} (M_*, z) \,,
\end{align} 
where ${\Phi_{\rm GSMF}(M_*, z)}$  is the GSMF which is given by a (double) Schechter function as
\m \label{GSMF_params}
\!\!\!\!\!{\Phi_{\rm GSMF}} &\!\!\!\!\!=\!\!\!\!\!& 
\text{ln}10 \ \exp\left(-10^{\text{log}_{10} \left(M_* - \mathcal{M}_{\star}  \right)} \right) \nonumber \\
&\!\!\!\!\!\times\!\!\!\!\!&\! \left( \phi_1 10^{ \left(\text{log}_{10} \left( M_* -   \mathcal{M}_{\star}  \right) \right) \left(a_1 + 1\right) }\!+\! \phi_2 10^{ \left(\text{log}_{10} \left( M_* -   \mathcal{M}_{\star}  \right) \right) \left(a_2 + 1 \right) } \right) \,. \nonumber \\
\n
Hereafter, we collectively denote the parameters $\{\mathcal{M}{\star}, \phi_1, a_1, \phi_2, a_2\}$ in the GSMFs as ${\Lambda}_{\text{GSMF}}$. 

\section{SGWB from SMBH binary population}
\label{sec_3}
The characteristic strain spectrum, $\hc^{2} (f)$ , from a cosmic population of SMBH binaries emitting GWs within a frequency bin $d\fgearth$, as observed on Earth, is given by:
\be
	\hc^{2} (\fgearth) = \fgearth \int \int \int dz ~dM_{\bullet} ~dq_{\bullet} ~\hs^{2} ~\frac{d^{4}N}{dz ~dM_{\bullet} ~dq_{\bullet} ~d\fgearth},
	\label{eqn:SGWBStrain}
\ee
where $d^{4}N$ represents the number of SMBH binaries within a given redshift range $dz$, primary black hole mass range $dM_{\bullet}$, and mass ratio range $dq_{\bullet}$, which are emitting GWs within a frequency range $df$. 
Additionally, $\hs$ denotes the polarization- and sky-averaged GW strain contributed by each individual source, which writes, 
\be
    \hs = \sqrt{\frac{32}{5}} \left(\frac{G M_{c}}{c^{3}} \right)^{5/3} \frac{\left( \pi \fgrest \right)^{2/3} c}{\dist}~,
    \label{eqn:strain}
\ee
where $M_c = \mbh (q_{\bullet}^{3}/(1+q_{\bullet}))^{1/5}$ is the chirp mass of the binary, $\dist$ is the proper (co-moving) distance to the binary, and $\fgrest$ is the frequency of the GWs emitted in the rest frame of the binary. The Earth-observed GW frequency $\fgearth$ is related to $\fgrest$ by  $\fgearth = \fgrest / (1+z)$.

We further rewrite the term $\frac{d^{4}N}{dz ~dM_{\bullet} ~dq_{\bullet} ~d\fgearth}$  in equation~(\ref{eqn:SGWBStrain}) as, 
\be
    \frac{d^{4} N}{dz ~dM_{\bullet} ~dq_{\bullet} ~d\fgearth} = \frac{d^{3} n_{\bullet}}{dz ~dM_{\bullet} ~dq_{\bullet}} ~\frac{dV}{dz} ~\frac{dz}{dt} ~\frac{dt}{d\fgearth} \,.
	\label{eqn:d4N_BH}
\ee
The conversion above transforms the number of binaries per co-moving volume element, $dV$, into the number of binaries per GW frequency bin, $d\fgearth$, by first converting to redshift, and then to the Earth-observed time. 
Once a binary hardens and decouples from its surrounding galactic environment, the evolution of its orbit becomes dominated by the emission of gravitational radiation, occurring at a rate given by:
\be
    \frac{df_{\rm orb}}{dt} = \frac{96}{5} \left( \frac{G M_{c}}{c^{3}} \right)^{5/3} ( 2 \pi )^{8/3} f_{\rm orb}^{11/3} \,, 
    \label{eqn:dfdt}
\ee
where $f_{\rm orb}$ is the orbital frequency in the rest frame. For binaries in a circular orbit, the frequency of GWs emitted in the rest frame is given by  $\fgrest = 2 f_{\rm orb}$. 

The frequency dependence of $\hc$  is encoded in both $\hs$ and $dt/df$.
By combining Eqs. \ref{eqn:SGWBStrain}-\ref{eqn:dfdt}, $\hc(f)$ can be expressed as a simple power-law with a dimensionless amplitude $\ayr$ referenced to a characteristic frequency of $\fyr = 1$ yr$^{-1}$,
\be
    h_{c} (f) = \ayr \left( \frac{\fgearth}{\fyr} \right)^{-2/3}.
\ee

By substituting equation~(\ref{relate_BH_merge_galaxy_merger}) into equation (\ref{eqn:d4N_BH}), and then into equation~(\ref{eqn:SGWBStrain}), and integrating over the mass ratio for major mergers, the strain of SGWB is finally given by: 
\begin{equation}
 \hc^{2} = f \!\! \int \!\!\! \int \!\!\! \int \!\!\! {\Phi_{\rm GSMF}(M_*, z')} \frac{dN}{dt}(M_*, z') \frac{dV}{dz} \left( \frac{dt}{df} h_{s}^{2} \right) dz~dM_* \,,
    \label{eqn:hc_gsmf}
\end{equation}
where $z'=z(t_L(z) + \tau)$ represents the redshift at which the galaxy merger occurs, 
while $z$ denotes the redshift at which the SMBH binary emits GWs in the PTA frequency band.   

\begin{figure}
\centering 
\includegraphics[width=0.35\textwidth]{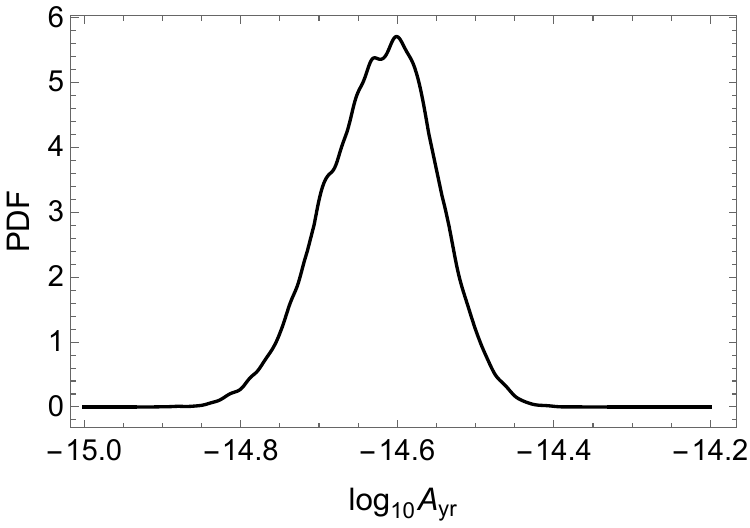}  
\caption{Strain amplitude of SGWB detected by NANOGrav \citep[][]{NANOGrav_2023}.
} 
\label{Figure.hc} 
\end{figure}

The strain amplitude of SGWB observed by NANOGrav \citep[][]{NANOGrav_2023} assuming quasi-circular orbit for SMBH binary is shown in Fig.~\ref{Figure.hc}. 
Similar strain amplitudes have been reported by other PTAs \citep[][]{EPTA:2023sfo, EPTA:2023fyk, Reardon_2023, CPTA_2023}. We use this strain amplitude to constrain the merger rates of SMBHs and their host galaxies in Section \ref{sec_5}. 

\section{Detecting SMBH binary merger events with LISA}
\label{sec_4}
LISA is expected to detect between a few to several hundred SMBH binary mergers per year \citep[e.g.][]{Sesana2007, Klein_2016}, depending on the redshift and mass range. The detection rate is highly sensitive to the assumptions about galaxy formation, SMBH growth, and the efficiency of binary mergers \citep[e.g.][]{Volonteri:2002vz, Sesana2007, Barausse:2012fy, Klein_2016}. 
In this section, we demonstrate how the merger rate of SMBH binaries can be inferred from LISA GW events. The mass of the merging SMBHs is considered in the range $\[10^{5}\Msun,10^{8}\Msun\]$. The lower limit reflects the smallest SMBH masses observed in galactic centers, while the upper limit corresponds to LISA's detection threshold, as determined by its signal-to-noise ratio curve.

\subsection{Population Analysis Framework}
\label{sec_frame}
The hierarchical Bayesian approach is a widely used method for inferring the underlying distribution of a population based on a set of observed events that statistically follow it. 
This distribution is typically described by a theoretical model parameterized by a few  hyperparameters. 
The hyperparameters are then constrained using observed data through Bayesian inference. This approach has been frequently applied to estimate the population properties of stellar-mass black hole binaries using GW events detected by LIGO/Virgo, such as those reported in the GWTC catalogs \citep[][]{Abbott_2021_GWTC2,LIGO_2021_GWTC3}. 

In this subsection, we outline the process of estimating the merger rate of SMBHs and galaxies using hierarchical Bayesian inference.  
 Given the LISA data ${\{\bf d\}}$ of SMBH merger events, the total number of merger events is modeled as an inhomogeneous Poisson process. The corresponding likelihood function is expressed as: 
\m \label{def_hyper_likelihood}
\mathcal{L}(\{{\bf d}\} | {\bm \Lambda} ) &\propto& {N_{\text{mod}}({\bf \Lambda})}^{N_{\text{det}}} e^{- N_{\text{exp}} ({\bf \Lambda}) }  \times  \\ \nonumber
&& \prod^{N_{\text{det}}}_{i=1} \int \mathcal{L}(d_i |{\bm \theta}) P ({\bm \theta}| {\bm \Lambda}) d {\bm \theta} \,,
\n 
 where $N_{\rm det}$ is the number of events detected during an observational period $T_{\text{det}}$, ${\bm \theta}=(m_1, m_2, z_1, z_2, ...)$ denotes source parameters, ${\bm \Lambda}$ refers to the hyperparameters of the merger rate model. $N_{\text{mod}}({\bf \Lambda})$ is the total number of events predicted by the model  ${\bf \Lambda}$ , while $N_{\text{exp}}({\bf \Lambda})=\xi({\bf \Lambda})  N_{\text{mod}}$ represents the expected number of events that could be detected, assuming a detection fraction $\xi({\bf \Lambda})$. 
 $\mathcal{L}(d_i |{\bm \theta})$ is the likelihood of an individual GW event $d_i$, and $P ({\bm \theta}| {\bm \Lambda})$ denotes the population distribution corresponding to model ${\bm \Lambda}$. 
 In this context, the detection fraction $\xi({\bm \Lambda})$ is assumed to be unity, as the SMBH binary merger events in the considered mass and redshift ranges typically have large signal to noise ratios \citep{2017eLISA}. 

The integration over ${\bm \theta}$ in the likelihood function of equation~(\ref{def_hyper_likelihood}) can be evaluated by averaging the population distribution $P ({\bm \theta}| {\bm \Lambda})$ over Monte-Carlo (MC) samples of ${\bm \theta_i}$, which are drawn from the likelihood function $\mathcal{L}(d_i |{\bm \theta})$ for each individual event $i$. This allows the expression to be rewritten as:
\m \label{redef_likelihood}
\mathcal{L}(\{{\bf d}\}|{\bm \Lambda}) \propto  
{N({\bf \Lambda})}^{N_{\text{det}}} e^{- N_{\text{exp}} ({\bf \Lambda}) } 
\prod^{N_{\text{det}}}_{i=1} \langle {P ({\bm \theta}| {\bm \Lambda}) \over P_{\varnothing} ({\bm \theta})} \rangle \,,
\n
where $ \langle ... \rangle$ is the average over samples of ${\bm \theta}_i$, and $P_{\varnothing} ({\bm \theta})$ is the default prior taken in the parameter estimation which is usually set to a uniform distribution. 

The posterior of the hyperparameters, $P ({\bm \Lambda} | \{ {\bf d} \})$, given the data $\{{\bf d}\}$, is the multiply of the likelihood $\mathcal{L}(\{{\bf d}\}|{\bm \Lambda})$ and prior $P ({\bm \Lambda})$ of the model   
\m
P ({\bm \Lambda} | \{ {\bf d} \}) \propto \mathcal{L}(\{{\bf d}\}|{\bm \Lambda}) P ({\bm \Lambda})\,.
\n
Here, we assume a uniform distribution for model prior $P ({\bm \Lambda})$ .

We consider a population model that describes the joint distribution of two key population properties: the total mass $M_{\bullet}$ of the SMBH binary and the redshift 
$z$ at their merger, i.e., ${\bm \theta}=(M_{\bullet},z)$. The corresponding population distribution, $P({\bm \theta}| {\bm \Lambda})$, is given by the normalized merger rate
$\frac{d^{2}n_{\bullet}}{dz ~d\text{log}_{10}M_{\bullet}}$ of SMBH binaries (equation~(\ref{BH_merger_rate})), which is defined as:
\m \label{population_prob}
P({\bm \theta}| {\bm \Lambda}) =  {1\over \mathcal{N}({\bf \Lambda})}\frac{d^{2}n_{\bullet}}{dz ~d\text{log}_{10}M_{\bullet}} ( M_{\bullet}, z | {\bf \Lambda}) \,,
\n
where 
\m
\mathcal{N}({\bf \Lambda})=\int\int \frac{d^{2}n_{\bullet}}{dz ~dM_{\bullet}} ( M_{\bullet}, z | {\bf \Lambda}){\rm d}  M_{\bullet} \ \text{d}z \,,
\n
is the number of merger events per year (in Earth time). The predicted total number of mergers, $N({\bf \Lambda})$, during an observational period $T_{\text{det}}$, is given by $N({\bf \Lambda})= \mathcal{N}({\bf \Lambda}) \ T_{\text{det}}$. 

\section{Inferring the merger rate of SMBHs and galaxies}
\label{sec_5}
In this section we generate mock LISA GW data from different cases of models, and 
estimate the merger rate of galaxies and SMBHs from the current PTA data and the mock LISA data. We then compare the reconstructed results with observations and theoretical simulations. 

\subsection{Inferring the Merger Rate of Galaxies}
\label{sec:5.1}
\subsubsection{Theoretical Models of Galaxy Merger Rates}
\label{sec:theoretical_merger_rate}
The  model of SMBH merger rate is determined by equation~(\ref{BH_merger_rate}). 
We further define the galaxy merger rate per galaxy per unit time as 
\begin{equation} \label{galaxy_merger_rate_model}
{dN\over dt_{\rm r}}(z,M_*|{\bm \Lambda}_{\text{dN/dt}}) = n_0 \left({M_*\over 10^{11}\Msun}\right) ^{\alpha_0 + \alpha_1 (1+z)} (1+z)^{\beta} \,,
\end{equation}
where $t_{\rm r}$ is the time measured at the source's rest frame. Then
\m  \label{model_rate_per_galaxy}
\mathcal{R}_{\rm Gal} (z,M_*|{\bm \Lambda}_{\text{dN/dt}}) &=& {dN\over dt}(z,M_*|{\bm \Lambda}_{\text{dN/dt}}) \\ \nonumber
&=& {1\over (1+z)}{dN\over dt_{\rm r}}(z,M_*|{\bm \Lambda}_{\text{dN/dt}}) \,,
\n
is the galaxy merger rate per galaxy per unit time, with $t$ denotes the time measured by earth observer. 

We consider two different models for ${dN\over dt_{\rm r}}$ by setting the model parameters in the following ways:\\
{\bf Case 1}: where ${\bm \Lambda}_{\text{dN/dt}} = \{ n_0, \alpha_{0}, \beta\}$, and $\alpha_{1}=0$. This model is commonly adopted in studies of the PTA SGWB strain generated by SMBH binary populations \citep[e. g., ][]{Chen:2018znx, EPTA:2023xxk}. \\
{\bf Case 2}: where ${\bm \Lambda}_{\text{dN/dt}} = \{ \alpha_{0}, \alpha_{1}, \beta\}$, and $n_0=0.03$. In this case, the merger rate $dN/ dt_{\rm r}$ with $\alpha_0 = 0.2$, $\alpha_1 = - 0.01$, and $\beta = 2.4$ serves as a simplified approximation to the best-fit results predicted by the Illustris simulation \citep[][]{Vicente2015} (see Fig.~\ref{Figure.dN_dtr} for details). It is worth noting that for $M_*<10^{11}\Msun$, a positive value of $\alpha_1$ acts as an exponential suppression of ${dN / dt_{\rm r}}$ at high redshift, as demonstrated by \citep[][]{DuanQiao2024}, where the authors fit the pair fraction and galaxy merger rate using a power-law + exponential model. 

\subsubsection{Relating the Merger Rate of Galaxies to the Merger Rate of SMBHs}
\label{calculation_seting}

By substituting equation~(\ref{model_rate_per_galaxy}) and (\ref{galaxy_rate2}) into equation~(\ref{BH_merger_rate}), the merger rate of SMBH binaries can be determined for a given population model ${\bm \Lambda}$.   
For $z<4$, we adopt the scaling relationship (equation~(\ref{scaling_relation})) based on local observations \cite[][]{Kormendy:2013dxa}, with $a = \text{ log}_{10}\kappa + 9 - 11* b$, $b =  1.17\pm 0.08$, and $\epsilon = 0.28$, where $\kappa=0.49 \substack{+0.06 \\ -0.05}$. 
For $z\ge4$, we adopt the scaling relationship based on recent JWST results
 \citep[][]{Pacucci_2024} with $a=-2.43\pm0.83$, $b=1.06\pm0.09$, and $\epsilon=0.69$. 
The GSMF (equation~(\ref{GSMF_params})) is derived from a series of observations spanning $z=0 -10.5$ \citep[][]{Baldry2012, HuertasCompany2016, Santini2012, McLeod2021, Song2016, Stefanon2021}. 
To consider the observational uncertainties of GSMF and $M_{\bullet}-M_*$ relation in the estimation of model parameters, the parameters ${\Lambda}_{\text{GSMF}}$ and $\Lambda_{M_{\bullet}-M_{*}}$ used in the Likelihood function for parameter estimation are randomly sampled from the parameter space constrained at each observational redshift bin.

In the following sections we assume a unit SMBH occupation fraction for galaxies. Since the lowest mass end of SMBHs we consider in this work is $M_{\bullet}=10^5\Msun$, the corresponding lower mass bound of galaxies in the scaling relationship is $M_*=10^7\Msun$. This unit occupation fraction assumption represents a rough approximation to the recent multiwavelength constraints reported by \cite{Burke_2025}, who find a local black hole occupation fraction of at least 90 per cent at a stellar mass of $M_* =10^8\Msun$ and at least 39 per cent at $M_* =10^7\Msun$. We also discuss a varying occupation fraction and the constrains in section \ref{sec_occupation}. 

\begin{figure*} 
\centering 
\includegraphics[width=0.8\textwidth]{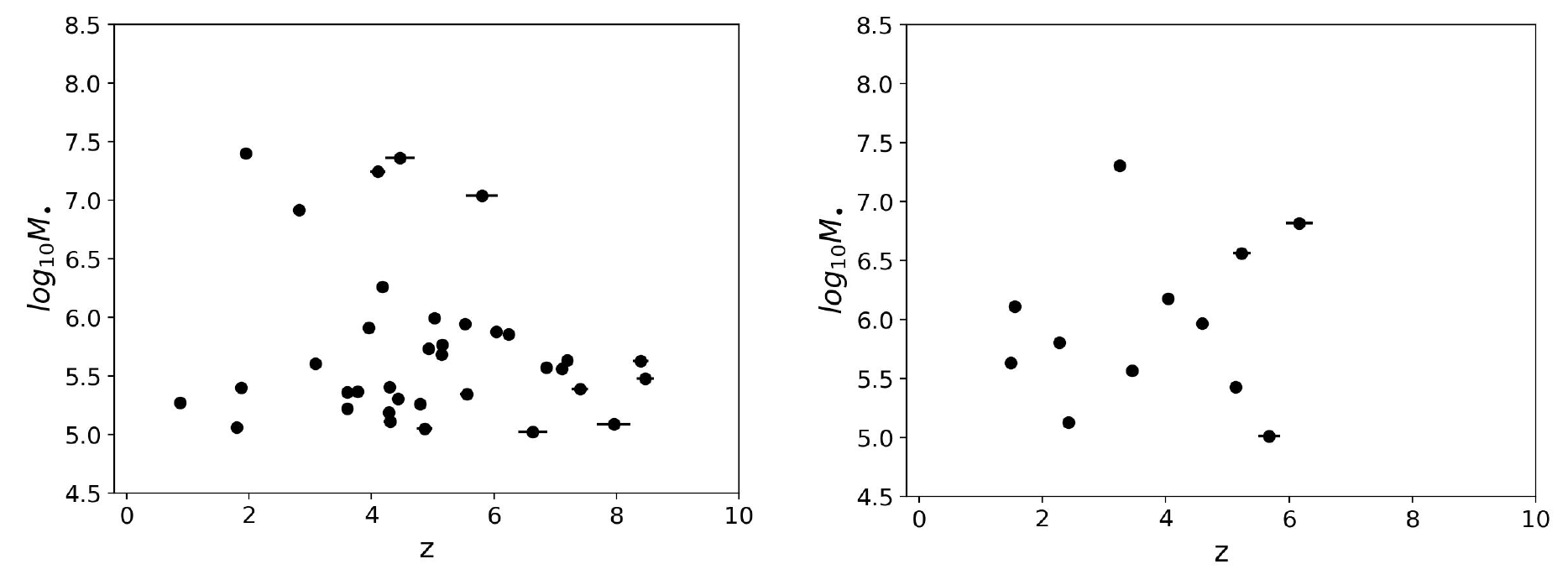}  
\caption{Left: the 36 mock LISA GW events from the first realization. Right: the 12 mock LISA GW events from the second realization. 
} 
\label{Figure.LISA_data} 
\end{figure*}
\begin{figure*}
\centering 
\includegraphics[width=0.8\textwidth]{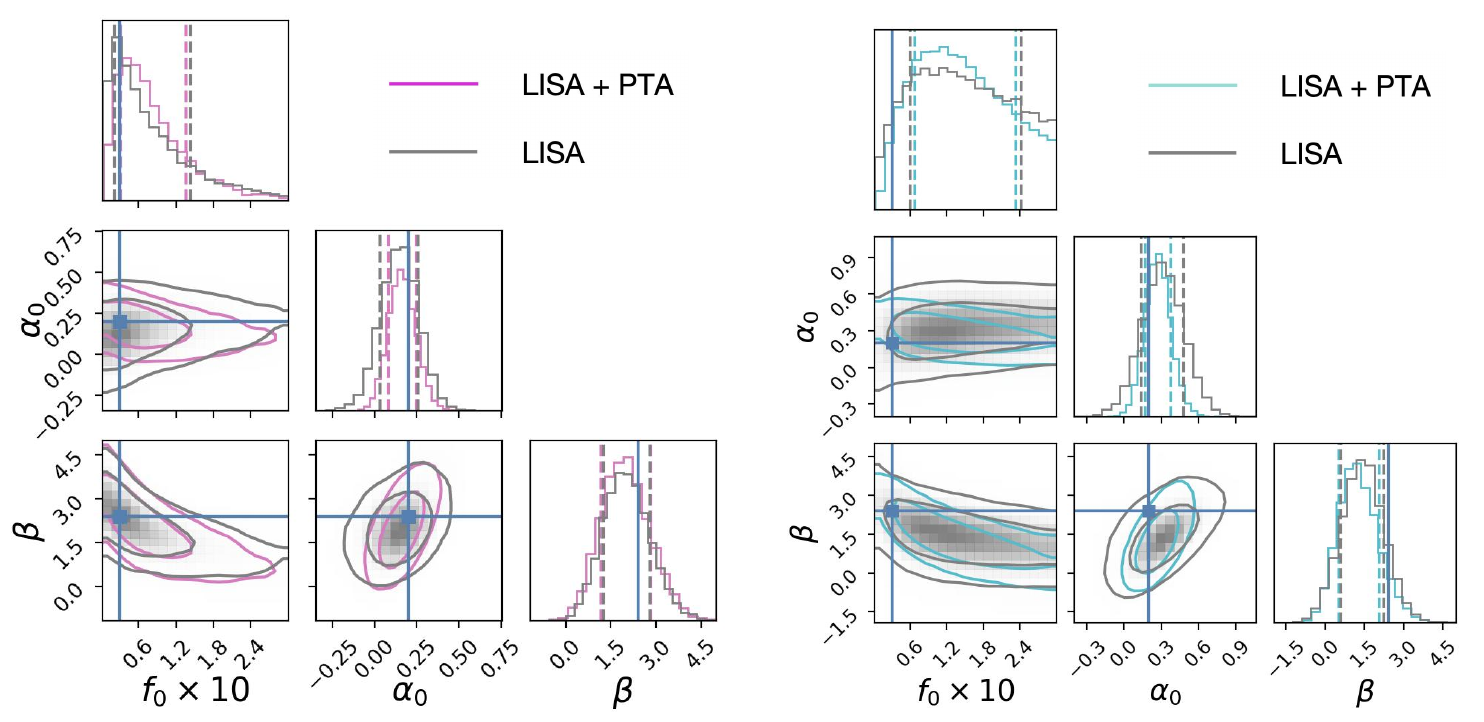}  
\caption{Left: posteriors of the model parameters ${\bf \Lambda}_{\text{dN/dt}}$ in Case 1 model of galaxy merger rate (discussed in subsection \ref{sec:theoretical_merger_rate}), estimated with mock LISA data from the first realization (left panel of Fig.~\ref{Figure.LISA_data}). The pink lines show the result constrained with both LISA mock data and current PTA detection of SGWB (Fig.~\ref{Figure.hc} ). The gray lines show the result inferred only with LISA mock data. Right: Similar to the left figure, but for the mock LISA data from the second realization (right panel of Fig.~\ref{Figure.LISA_data}). In both panels, the horizontal and vertical blue lines represent the injected values.
} 
\label{Figure.posteriors_case1_eg1_eg2} 
\end{figure*}
\begin{figure*} 
\centering 
\includegraphics[width=0.8\textwidth]{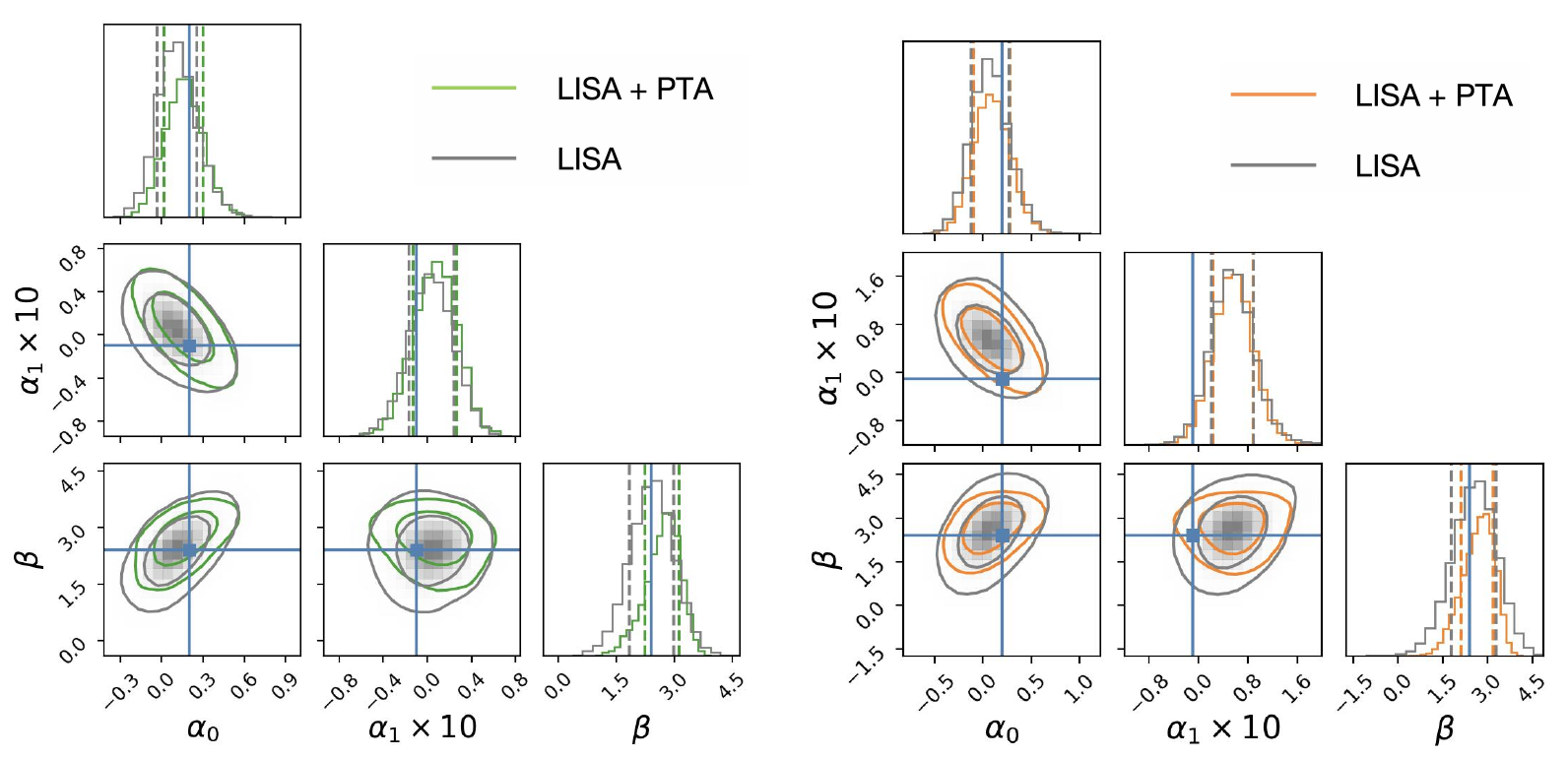}   
\caption{
Similar to Fig.~\ref{Figure.posteriors_case1_eg1_eg2}, but for the Case 2 model of galaxy merger rate discussed in subsection \ref{sec:theoretical_merger_rate}. 
} 
\label{Figure.posteriors_case2_eg1_eg2} 
\end{figure*}

\subsubsection{Estimating Galaxy Merger Rates Using Mock LISA Data and PTA Observations}
\label{infer_merger_rate}
Here, we present two realizations of LISA GW events in Fig~\ref{Figure.LISA_data}, which displays the mass–redshift distribution of the these events. In both cases, the SMBH binary merger rate is computed using a taken galaxy merger rate model along with specific GSMFs and $M_{\bullet}-M_*$ relations. The galaxy merger rate per galaxy adopts the same set of model parameters with $n_0 = 0.03$, $\alpha_0 = 0.2$, $\alpha_1 = - 0.01$, and $\beta = 2.4$ (see equation~(\ref{galaxy_merger_rate_model})). 
These parameters correspond to the injected values used to generate the mock datasets, which will later be used to constrain the (initially unknown) model parameters.   
The comparison between this assumption of galaxy merger rate and the ones predicted from cosmological simulations or semi-analytical models \citep[][]{Vicente2015, OLeary_2021, Filip_2022_galaxy} is illustrated in Fig.~\ref{Figure.dN_dtr}. 
The parameters $\Lambda_{\text{GSMF}}$ and $\Lambda_{M_{\bullet}-M_{*}}$ in the GSMFs and $M_{\bullet}-M_*$ relations are randomly sampled from observational constraints across different redshift bins. The corresponding parameter sets for the two realizations, labelled GW(eg1) and GW(eg2), are provided in Tables \ref{tab:GSMF} and \ref{tab:scaling_relation}, respectively. 
The GW events shown in Fig.~\ref{Figure.LISA_data} are then sampled from the corresponding SMBH binary merger rate.
 The error bars of mass and redshift in each event is calculated using the Fisher information matrix with the phenomenological waveform model PhenomA \citep{Ajith_2007, Ajith_etal_2011}. The small error bars reflect LISA’s high signal to noise ratio for these events. 
The difference of the event number and merger distribution between these two models is due to the different sets of $\Lambda_{\text{GSMF}}$ and $\Lambda_{M_{\bullet}-M_{*}}$ from the sampling, rather than the different choices of model parameters, e.g., ${\bm \Lambda}={\bm \Lambda}_{\text{dN/dt}}$. 
For the current discussion, no delay time between galaxy mergers and SMBH mergers is assumed. The non-vanishing delay time is discussed in subsections \ref{sec_delay_time}-\ref{sec_occupation}. 

Figs.~\ref{Figure.posteriors_case1_eg1_eg2} and \ref{Figure.posteriors_case2_eg1_eg2} present the posterior distributions for the model parameters ${\bf \Lambda}_{\text{dN/dt}} = \{ n_0, \alpha_{0}, \beta \}$ in Case 1 and ${\bf \Lambda}_{\text{dN/dt}} = \{ \alpha_{0}, \alpha_{1}, \beta \}$ in Case 2, as estimated from the two realizations of mock data shown in  Fig~\ref{Figure.LISA_data}. 
In each estimation process, we compare the results inferred from the joint (mock) LISA + PTA data (red lines) with those inferred solely from the mock LISA data (gray lines). 
From the figures, we can see that:\\
a. LISA's detection of SMBH binary merger events plays a pivotal role in recovering the merger rates of galaxies and SMBHs. Additionally, incorporating PTA constraints on the SGWB strain further enhances the precision of model parameter estimation. This is expected, as LISA-like detectors can effectively measure the merger distribution, such as in the joint mass-redshift distribution discussed here, and fit the data using a hierarchical Bayesian inference approach;  \\
b. A larger number of detected events yields tighter constraints on the model parameters. 
For the model parameters estimated from the first GW mock dataset (left panel of Fig.~\ref{Figure.LISA_data}), which contains more events, the injected truth values lie well within the corresponding $1\sigma$ credible regions (left panels of Figs.~\ref{Figure.posteriors_case1_eg1_eg2} and \ref{Figure.posteriors_case2_eg1_eg2}). In contrast, for the model parameters  estimated from the second mock GW data (right panel of Fig.~\ref{Figure.LISA_data}), which contains fewer events, the injected truth values ($n_0$, $\alpha_1$, and $\beta$) deviate from the $1\sigma$ credible regions (right panels of Figs.~\ref{Figure.posteriors_case1_eg1_eg2} and \ref{Figure.posteriors_case2_eg1_eg2}). The deviations of these parameters from their peaks is attributed to both systematic biases—arising from observational uncertainties in the adopted GSMFs and $M_{\bullet}-M_*$ relation—and random biases introduced when generating the mock LISA data from theoretical merger rates. The latter may be mitigated if LISA detects a larger number of GW events, as a larger statistical sample more accurately reflects the true merger rate, reducing uncertainties in parameter estimation; \\
c. The model parameters $\{ \alpha_{0}, \alpha_{1}, \beta\}$ in the Case 2 model are well constrained, whereas the parameter $n_0$ in the Case 1 model exhibits significant degeneracy, particularly in the second realization, which includes fewer GW events. 
\begin{figure*} 
\centering 
\includegraphics[width=0.95\textwidth]{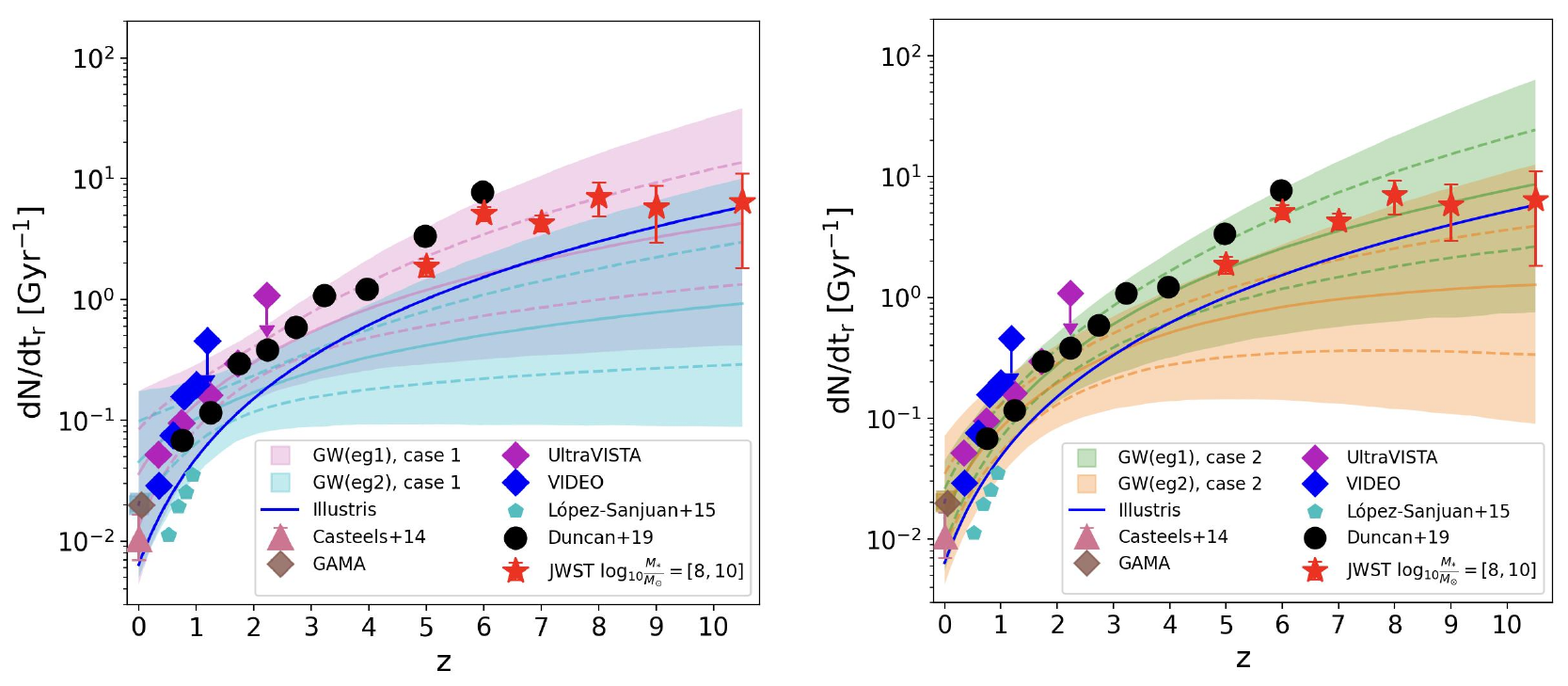}  
\caption{
The galaxy merger rate ${dN\over dt_{\rm r}}(z) $ for galaxy stellar mass in the range $\text{log}_{10} {M_* }/\Msun = \[8,10\]$. 
Left panel: The pink (cyan) lines represent the recovered merger rate, ${dN\over dt_{\rm r}}(z)$, based on the posterior distributions of model parameters presented in the left (right) panel of Fig.~\ref{Figure.posteriors_case1_eg1_eg2}, using the LISA + PTA data.
Right panel: The green (orange) lines represent the recovered merger rate, ${dN\over dt_{\rm r}}(z)$, based on the posterior distributions of model parameters presented in the left (right) panel of Fig.~\ref{Figure.posteriors_case2_eg1_eg2}, using the LISA + PTA data. 
In both panels, the filled regions represent the $2-\sigma$ confidence interval, the dashed lines represent the corresponding $1-\sigma$ confidence interval, and the solid (pink and cyan) lines represent the medium values. 
The shaped points with error bars represent the galaxy merger rates derived from the observations of galaxy pairs across different redshift bins \citep[][]{Casteels2014, Sanjuan2015, Duncan_2019, Conselice_2022} , in particular, the red stars show the results from recent JWST observations \citep[][]{DuanQiao2024}. We also show the merger rate for galaxy stellar mass in the range $\text{log}_{10} {M_* }/\Msun = \[8,10\]$ predicted by the Illustris simulation \citep[][]{Vicente2015, OLeary_2021, Filip_2022_galaxy} in the blue lines. 
} 
\label{Figure.recover_galaxy_rate_eg1_eg2} 
\end{figure*}

\subsubsection{Comparing Galaxy Merger Rates Inferred from GWs with Those Derived from Galaxy Pair Observations}
\label{sec:compare_merger_rate}
The galaxy merger rate, ${dN\over dt}(z)$, could also be observationally derived from studies of galaxy pairs across different redshifts \citep[e.g., ][]{Casteels2014, Sanjuan2015, Duncan_2019, Conselice_2022, DuanQiao2024, Puskas_2025} or theoretically predicted using cosmological simulations or semi-analytical models \citep[e.g., ][]{Vicente2015, OLeary_2021, Filip_2022_galaxy}. 
 This rate is directly related to the galaxy pair fraction through the following way:
\m \label{pair_fraction}
 {dN\over dt}(z) = {f_{\text{p}}(z, M_{*}) \over \text{T} (z)} \,,
\n
where $f_{\text{p}}$ is the galaxy pair fraction, and $\text{T}$ is the merger timescale of the galaxy pairs. 

Recently,  \cite{DuanQiao2024} conducted a JWST study on galaxy pair fractions and galaxy merger rates, extending the constraints up to $z=11.5$ \citep[][]{DuanQiao2024}. In their work, the authors found that the galaxy pair fraction, $f_{\text{p}}$, is better fit to the data using a power-law + exponential parameterization, expressed as:
\m
f_{\text{p}}= f_{\text{c}} \times (1+z)^{m} \times e^{\kappa (1+z)} \,. 
\n
The merger timescale $\text{T}$ for galaxy pairs in equation~(\ref{pair_fraction}) could be determined from cosmological simulations. 

In Fig.~\ref{Figure.recover_galaxy_rate_eg1_eg2}, we compare the galaxy merger rate, ${dN\over dt_{\text{r}}}(z)$, as a function of redshift for stellar masses in the range $\text{log}_{10} {M_* / \Msun} = \[8,10\]$, as recovered from GW detections (pink and cyan lines in the left panel, and green and orange lines in the right panel) , to results derived from observations of galaxy pairs \citep[][]{Casteels2014, Sanjuan2015, Duncan_2019, Conselice_2022, DuanQiao2024}  (represented by shaped points). Additionally, we compare these results to predictions from the Illustris simulation \citep[][]{Vicente2015} (blue lines). 

From Fig.~\ref{Figure.recover_galaxy_rate_eg1_eg2}, we can conclude that: \\
a. The two different realizations of LISA GW data (Fig.~\ref{Figure.LISA_data}) result in distinct estimates of the galaxy merger rate for both Case 1 (pink and cyan lines) and Case 2 (green and orange lines) models.
The differences between the two cases become more obvious at higher redshifts, though the associated error bars also increase.
Specifically, the galaxy merger rate recovered using the LISA data from the first realization (left panel of Fig.~\ref{Figure.LISA_data}) match better with the rates derived from galaxy pair observations for both Case 1 (pink lines) and Case 2 (green lines) models , though the Case 2 model match the data slightly better compared to the Case 1 model in this example.   
In contrast, the results obtained from the second realization of LISA data (right panel of Fig.~\ref{Figure.LISA_data}) for both cases (cyan and orange lines) show a noticeable mismatch with the observationally derived rates at higher redshifts. \\
b. For the same GW dataset, the reconstructed $2-\sigma$ confidence regions for the Case 1 and Case 2 merger rate models are largely consistent, except that the Case 1 model generally exhibits larger uncertainties and a systematic bias toward higher values at lower redshifts ($z < 2$), as well as a slight bias toward lower values at higher redshifts ($z > 3$).  
To test the robustness of the results, we have repeated the recovery of the galaxy merger rate, ${dN\over dt_{\text{r}}}(z)$, for both Case 1 and Case 2 models using different mock LISA datasets generated by varying the parameters $\{n_0, \alpha_0, \alpha_1, \beta\}$. The results, which are consistent with those shown in Fig.~\ref{Figure.recover_galaxy_rate_eg1_eg2}, suggest that the observed trends represent general characteristical behaviors of the Case 1 and Case 2 models. \\
c. The galaxy merger rate, ${dN/ dt}$, recovered from GW detections can be directly compared to those obtained from electromagnetic observations of galaxy pairs and theoretical simulations. These different and independent methods for deriving the galaxy merger rate serve as a valuable consistency check for one another. 

We also repeated the above analysis under the assumption that the $M_{\bullet}-M_*$ relation remains fixed to its local form \citep[][]{Kormendy:2013dxa}  across the redshifts. The resulting posteriors of the model parameters, estimated from the GW data shown in Figs.~\ref{Figure.hc}-\ref{Figure.LISA_data} are presented in Fig.~\ref{Figure.posteriors_local_relation}. The corresponding galaxy merger rate,  ${dN\over dt_{\text{r}}}(z)$, reconstructed from these posteriors, is shown in Fig.~\ref{Figure.recover_galaxy_merger_rate_local_relation}. 
Under this assumption, the galaxy merger rates inferred from the two GW datasets in Fig.~\ref{Figure.LISA_data} also diverge at high redshifts, similar to the behavior in Fig.~\ref{Figure.recover_galaxy_rate_eg1_eg2}, which assumes an $M_{\bullet}-M_*$ relation at $z > 4$ constrained by JWST observations \citep[][]{Pacucci_2024}. 
The primary difference is that Fig.~\ref{Figure.recover_galaxy_merger_rate_local_relation} shows systematically higher merger rates with smaller uncertainties at high redshifts compared to Fig.~\ref{Figure.recover_galaxy_rate_eg1_eg2}. 
The smaller error bars arise from the smaller uncertainties assumed in the $M_{\bullet}-M_*$ relation \citep[][]{Kormendy:2013dxa}. 
The systematically higher inferred merger rates stem from the fact that the local $M_{\bullet}-M_*$ relation, e.g. \cite{Kormendy:2013dxa}, typically predicts smaller SMBH masses compared to those implied by JWST observations \citep[][]{Pacucci_2024}. As a result, when mapping GSMFs to SMBH number densities via this scaling relation, the local relation yields lower SMBH number densities for a given mass. 
To reproduce the same GW event distribution, this scenario requires a higher merger rate per SMBH (or per galaxy) to compensate for the reduced number density. 
\subsection{Inferring the Delay Time Between Galaxy and SMBH Mergers}
\label{sec_delay_time}
After galaxy mergers, the SMBHs at the centers of the progenitor galaxies typically undergo an extended evolutionary timescale before coalescing. 
This delay is primarily governed by the timescale associated with the final parsec evolutionary phase  \citep[e.g.,][]{Milosavljevi__2001,Yu2002, Milosavljevic2003b}. 
Theoretically, SMBH binaries could merge efficiently via the interaction with environment,  e.g., if they reside in a triaxial-shaped stellar distribution environment \citep[e.g.,][]{Yu2002, Merritt_2004, Holley-Bockelmann:2006gbs, Gualandris2016}, a gas-rich environment \citep[e.g.,][]{Armitage2002, Dotti_2006_nucleardisc, Haiman_2009}, or they went through multiple mergers \citep[e.g., ][]{Hoffman_Loeb_2007}. 

\begin{figure}
\centering 
\includegraphics[width=0.95\columnwidth]{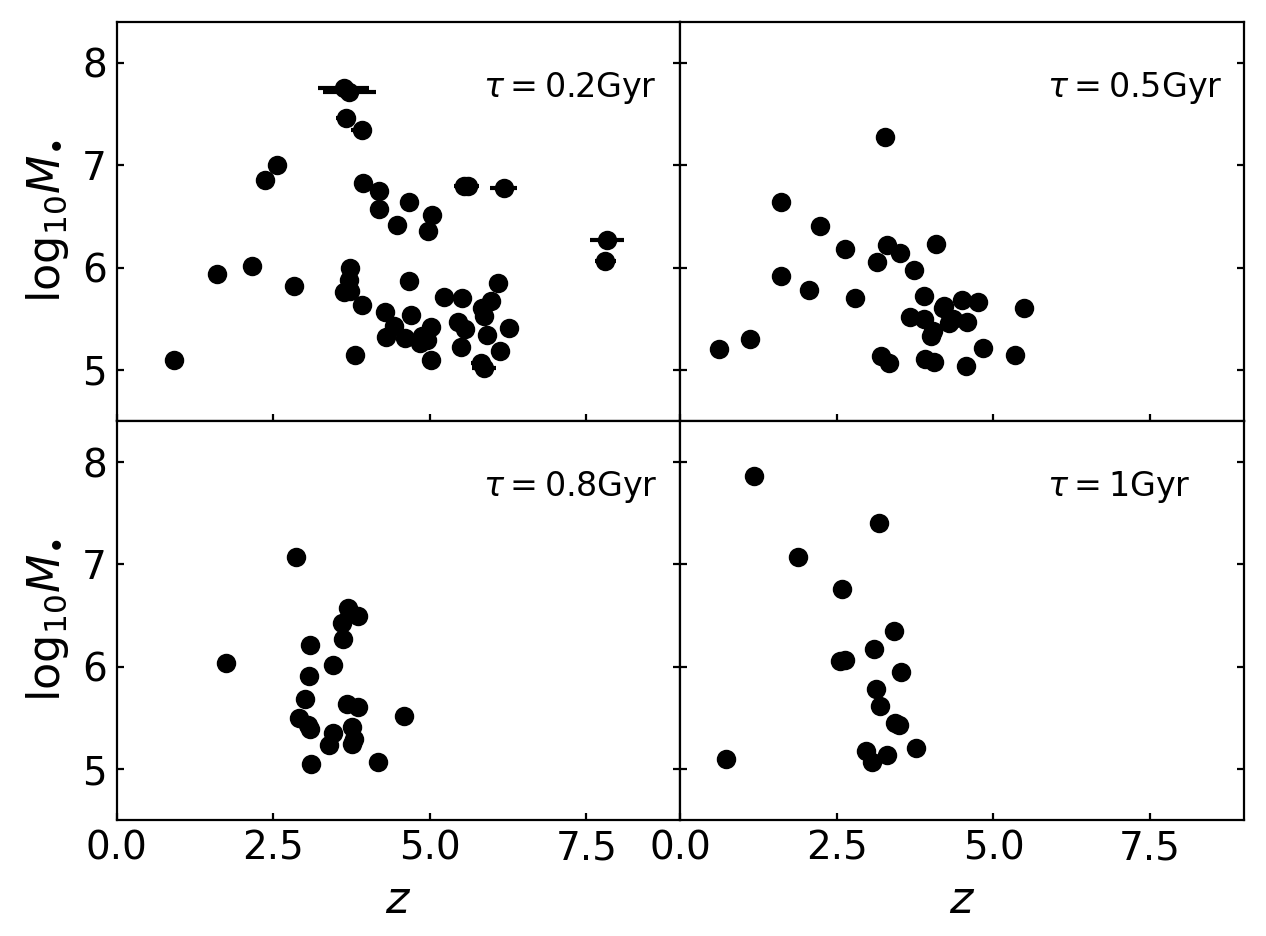}  
\caption{ Mock LISA GW data generated assuming a delay time of $\tau=0.2\text{Gyr}$, $0.5\text{Gyr}$, $0.8\text{Gyr}$, and $1 \text{Gyr}$ respectively. 
} 
\label{Figure.GW_data_delay} 
\end{figure}

In this subsection, we assess the feasibility of inferring the delay timescale between galaxy mergers and SMBH coalescences using the method proposed in this work. Following the approach described in subsection \ref{infer_merger_rate}, we generate mock LISA datasets based on a merger rate model with $\{n_0, \alpha_0, \alpha_1, \beta\} = \{0.03, 0.2, -0.01, 2.4\}$  (see equation~(\ref{galaxy_merger_rate_model})) , and assume fixed delay times of $\tau = 0.2,\mathrm{Gyr}$, $0.5,\mathrm{Gyr}$, $0.8,\mathrm{Gyr}$, and $1,\mathrm{Gyr}$ for comparison. These values represent characteristic timescales predicted in scenarios where SMBH binaries harden through interactions with stars in the loss cone of triaxial galaxies \citep[see, e. g., ][]{Yu2002,Khan_2011, Vasiliev_2015}. The GSMFs and $M_{\bullet}-M_*$ relations used at different redshift bins are randomly sampled from observational constraints and listed in Tables \ref{tab:GSMF}-\ref{tab:scaling_relation}. The four sets of mock LISA data generated under the above assumptions are shown in Fig.~\ref{Figure.GW_data_delay}, where the small error bars in mass and redshift reflect the high signal-to-noise ratios of these events.

\begin{figure*}
\centering 
\includegraphics[width=0.95\textwidth]{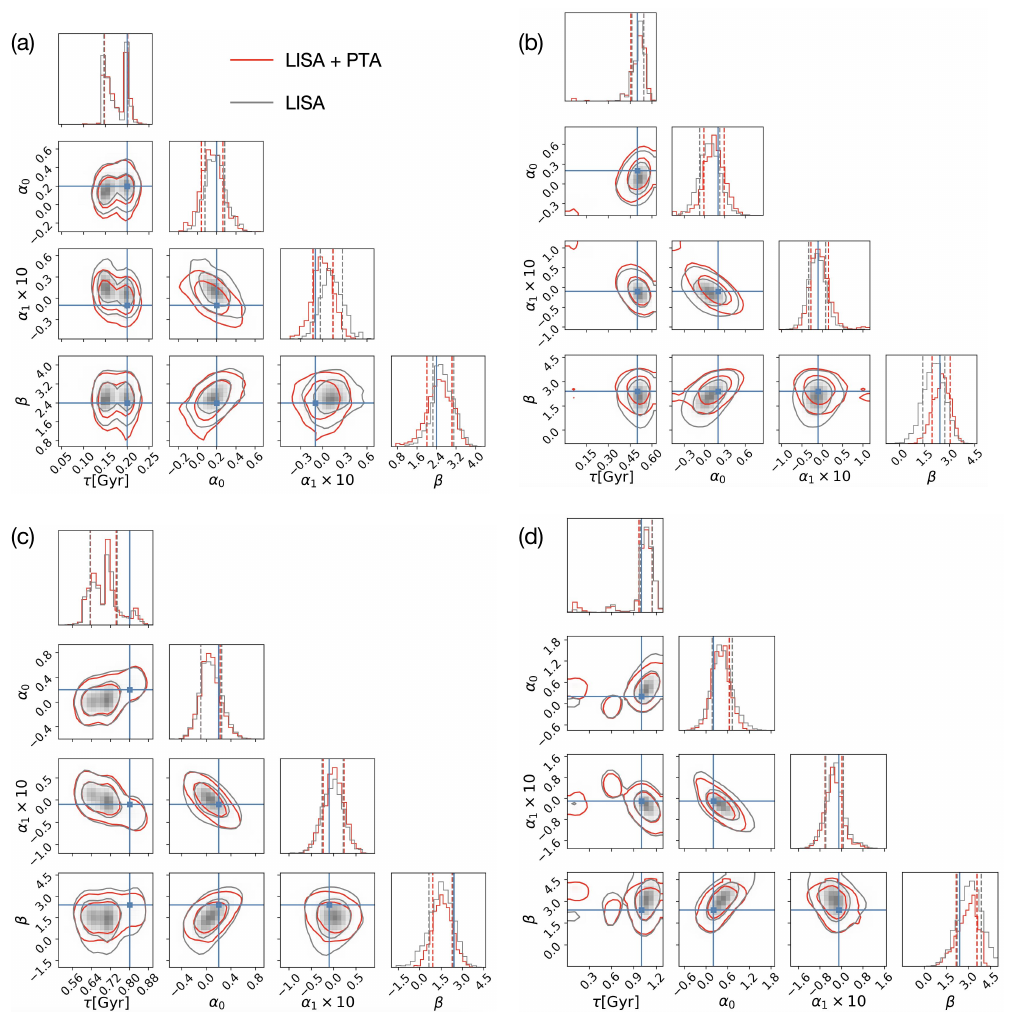}  
\caption{ Posterior distributions of the model parameters $\Lambda=\{\tau, \alpha_0, \alpha_1, \beta\}$, inferred from the mock LISA data presented in Fig.~\ref{Figure.GW_data_delay}. Subfigures (a) - (d) show the results for delay times of $\tau = 0.2 \text{Gyr}, 0.5 \text{Gyr}, 0.8 \text{Gyr}$ and $1 \text{Gyr}$, respectively. The horizontal and vertical blue lines represent the injected parameter values. 
} 
\label{Figure.posteriors_delay} 
\end{figure*}

The model parameters in this population model of SMBH binary merger rate (equation~(\ref{BH_merger_rate})) is now described by ${\bm \Lambda}=\{\tau, {\bf \Lambda}_{\text{dN/dt}}\}$, with ${\bf \Lambda}_{\text{dN/dt}} = \{ \alpha_0, \alpha_1, \beta \}$. 
The posteriors of the model parameters inferred from the four sets of mock LISA data (Fig.~\ref{Figure.GW_data_delay}) are shown in subfigures (a)-(d) of Fig.~\ref{Figure.posteriors_delay} respectively, where the gray lines represent the results inferred from the mock LISA data only, and the red lines correspond to the results obtained by jointly analysing these mock LISA data with the PTA constraints on the SGWB strain.  
As illustrated in Fig.~\ref{Figure.posteriors_delay}, the delay times in these four examples are properly constrained from the mock LISA data. Incorporating PTA constraints improves parameter estimation by either narrowing the $1\sigma$ credible regions or shifting their centres closer to the injected truth values (shown as vertical lines).  
For some model parameters, the injected values lie near the edges of the $1\sigma$ posterior distributions rather than at their peaks. Notably, the injected value $\tau = 0.8,\text{Gyr}$ in subfigure (c) of Fig.~\ref{Figure.posteriors_delay} falls outside the $1\sigma$ region, while still within the $2\sigma$ contour. These deviations can be attributed to both systematic biases—arising from observational uncertainties in the adopted GSMFs and $M_{\bullet}-M_*$ relation—and random biases introduced when generating the mock LISA data from theoretical merger rates. The latter may be mitigated if LISA detects a larger number of GW events, enabling the event distribution to better reflect the true merger rate. This could be alleviated either through a higher intrinsic merger rate in the LISA frequency band or by extending the mission duration beyond its nominal four-year baseline. The bimodal and subpeak structures observed in the posteriors of the delay time $\tau$ in subfigures (a) and (c) arise from a combination of model degeneracies and limited observational constraints.

In this work, we only consider a constant delay model due to the computational challenges of performing high-dimensional integrations as well as dealing with large model uncertainties (i.e., observational uncertainties from the GSMFs and scaling relations) in the likelihood. 
While a delay time model with a distribution would be more realistic, it is computationally expensive, prone to convergence issues, and exceeds our current resources. In our previous work \citep[][]{Fang:2022cso}, we investigated delay-time models with specific distributions while assuming fixed galaxy merger rate model. Future studies should explore distributed delay time models in the framework of this context to provide a more comprehensive understanding of SMBH binary evolution. 


\subsection{Mass Assembly of SMBHs Driven by Mergers}
\label{sec:SMBH_mass_assembly}
The merger rate of SMBH binaries, $\frac{d^{2}n_{\bullet}}{dz ~dM_{\bullet}}$, can be reconstructed by substituting the posterior distribution values of ${\bm \Lambda}$ into equation~(\ref{BH_merger_rate}). 
Using the reconstructed merger rate, one can further derive the mass assembly of SMBHs contributed by mergers, expressed as: 
\m \label{mass_assemble}
{d^2 M_{\bullet} \over dV dt} = \Phi_{\bullet}(M_{\bullet}, z) \mathcal{R}_{\bullet} (M_{\bullet}, z)M_{\bullet} q_{\bullet} \,, 
\n
where $M_{\bullet} q_{\bullet}$ is the mass of the secondary SMBH. Since we are considering major mergers, the mass ratio $q_{\bullet}\in \[1/4,1\]$. For simplicity, we assume $q_{\bullet}=0.6$, representing an average value for $q_{\bullet}$. 

\begin{figure*} 
\centering 
\includegraphics[width=0.9\textwidth]{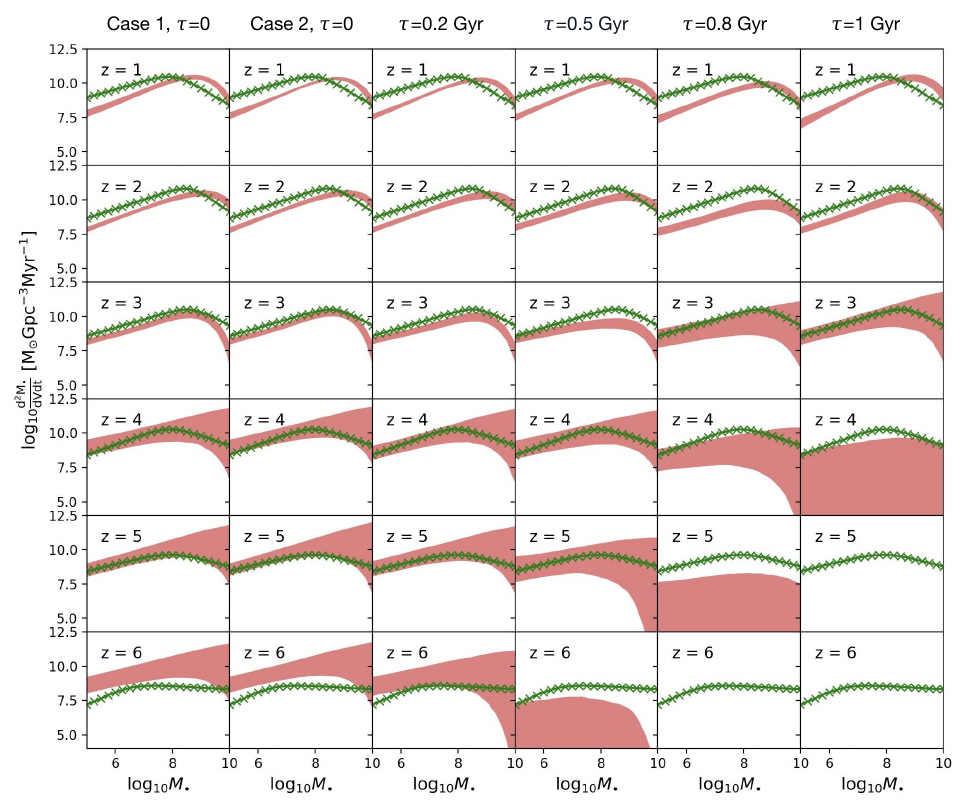}  
\caption{
The comparison of the contributions from mergers ( $1-\sigma$ credible region in the red shaded region, this work) and accretion (crossed green lines) to the mass assembly of SMBHs for different redshifts $(z=1-6)$ and SMBH mass bins $(\text{log}_{10} M_{\bullet}/\Msun=\[5,10\]$). 
First column: SMBH mass assembly via mergers, derived from the estimated galaxy merger rate assuming the Case 1 model with the first realization of GW data (pink lines and shaded region in Figs.~\ref{Figure.posteriors_case1_eg1_eg2} and \ref{Figure.recover_galaxy_rate_eg1_eg2}). 
Second column: Similar to the first column but based on the galaxy merger rate from the Case 2 model (red lines and shaded region in Figs.~\ref{Figure.posteriors_case2_eg1_eg2} and \ref{Figure.recover_galaxy_rate_eg1_eg2}). 
Third to sixth columns: Similar to the first two columns but based on the model assuming a delay time of $\tau = 0.2\text{Gyr}$, $0.5\text{Gyr}$, $0.8\text{Gyr}$, and $1\text{Gyr}$ respectively (red lines in subfigures (a)-(d) of Fig.~\ref{Figure.posteriors_delay}). 
The crossed green lines represent the SMBH mass assembly contributed by accretion, based on the simulation conducted by \citep{Pacucci_2020}. 
} 
\label{Figure.SMBH_mass_assemble} 
\end{figure*}

Fig.~\ref{Figure.SMBH_mass_assemble} illustrates the contribution of mergers (red shaded region, this work) and accretion (crossed green lines, from \cite{Pacucci_2020} ) to SMBH mass assembly. 
The first and second columns of Fig.~\ref{Figure.SMBH_mass_assemble} show that the galaxy merger rates, $dN/dt_\text{r}$, in the Case 1 (pink lines and shaded region in Figs.~\ref{Figure.posteriors_case1_eg1_eg2} and \ref{Figure.recover_galaxy_rate_eg1_eg2}) and Case 2 (green lines and shaded region in Figs.~\ref{Figure.posteriors_case2_eg1_eg2} and \ref{Figure.recover_galaxy_rate_eg1_eg2}) models produce a similar SMBH merger rate—and hence a similar SMBH mass assembly via mergers—at  $z>2$. 
However, at $z<2$, the Case 1 model is biased towards a higher SMBH merger rate compared to the Case 2 model, a trend also reflected in the recovered galaxy merger rates shown in Fig.~\ref{Figure.recover_galaxy_rate_eg1_eg2}. 
The merger mass assembly in the first two columns at $z\ge 4$ exhibits large uncertainties, due to the uncertainties in the SMBH mass function propagated from the scaling relationship and GSMFs, and the uncertainties in the estimated galaxy merger rate at this redshift bin.  

The delay time of SMBH binary mergers will have a negative impact on their merger rate/merger assembly, as illustrated by the comparison between the cases with delay (third to sixth columns) and without delay (first and second columns) in Fig.~\ref{Figure.SMBH_mass_assemble}.  The suppression becomes more pronounced with increasing delay time, leading to a steeper decline in the SMBH merger rate at high redshift (e.g., $z > 4$ in the last four columns).
In particular, the complete absence of mergers at $z\ge6$ and $z\ge5$ in the fifth and sixth columns, respectively, arises from the assumed delay times of $\tau = 0.8 \text{Gyr}$ and $1 \text{Gyr}$ in the corresponding models. 
 
By comparing the contributions to SMBH mass assembly from mergers and accretion, one can assess which process dominates SMBH growth across different redshifts. As shown in Fig.~\ref{Figure.SMBH_mass_assemble}, accretion generally dominants the mass assembly at $z<3$ for all the considered models, except for the most massive SMBHs ($\text{log}_{10}M_{\bullet}/\Msun> 8$ or $9$) at lower redshifts ($z \lesssim 1$–2). At higher redshifts ($z \gtrsim 4$–5), mergers become the dominant growth channel when the delay time is short ($\tau\le 0.2\text{Gyr}$), whereas accretion overtakes mergers for longer delays ($\tau \ge 0.5\text{Gyr}$). This trend is consistent with the findings of the BRAHMA simulation \citep[][]{Bhowmick2024}, which also shows that mergers dominate early SMBH growth when the merger delay timescale is short. 
 
  \subsection{SMBH Occupation Fraction of Galaxies}
\label{sec_occupation}
The occupation fraction refers to the proportion of galaxies that host a SMBH at their centers. It encodes key information about the formation and evolution of SMBHs and directly impacts the expected rate of SMBH mergers, influencing predictions for GW signals detectable by LISA. Observations in the local universe indicate that nearly all massive galaxies contain SMBHs \citep[e.g.,][]{Kormendy:2013dxa}, while the occupation fraction in low-mass dwarf galaxies remains uncertain \citep[e.g.,][]{Miller2015, Burke_2025, Greene2020}. Constraining the occupation fraction in dwarf galaxies is critical for distinguishing between different SMBH seed formation scenarios \citep[e.g.,][]{Natarajan2014, Inayoshi2020}. 

In the previous sections, we have assumed a unit occupation fraction (for those galaxies corresponding to SMBHs with masses above $10^5\Msun$ in the scaling relationship), which is a rough approximation to the recent multiwavelength constraints reported by \cite{Burke_2025}. To further investigate the potential of gravitational-wave detections to constrain the occupation fraction for galaxy with low stellar mass, we adopt a mass-dependent parameterization of occupation fraction $f_{\text{occ}}$ following equation~(3) of \cite{Beckmann2023} and \cite{Langen2025}, which writes, 
\m \label{occupation_fraction}
 f_{\text{occ}} = 1-{ f_i \over {1+\left( {\text{log}_{10} \left(M_*\right)\over {M^*}_{i}} \right)^{\epsilon_{i}} }} \,, 
\n
where the parameters $\{ f_i, {M^*}_{i}, \epsilon_{i} \}$ at $z<1$, $1 \leq z < 2$, and $2\leq z < 3$ are assigned the corresponding values at $z=0.25$, $1$, and $2$ respectively, as listed in Table 1 of \cite{Beckmann2023}. For $z\geq3$, we follow a similar treatment in \cite{Langen2025} by fixing ${M^*}_{3} = 8.55$ and $\epsilon_{3} = 33.75$ while keep $f_3$ as free parameter. The value of occupation fraction $f_{\text{occ}}$ at the low-mass end of $M_*$ at high redshift ($z\geq 3$) is  primarily determined by $f_3$. Constraining $f_3$ therefore plays an important role in probing the formation of SMBH seeds. 

We extend the set of model parameters ${\bf \Lambda}$ by including the occupation parameter $f_3$, such that ${\bf \Lambda}=\{\tau, f_3, {\bf \Lambda}_{\text{dN/dt}} \}$, with ${\bf \Lambda}_{\text{dN/dt}} =\{ \alpha_0, \alpha_1, \beta\}$, and jointly constrain them using mock LISA data in combination with PTA limits on the SGWB. Similar to the previous procedure, we generate two sets of mock LISA data assuming $\{f_3, \tau, n_0, \alpha_0, \alpha_1, \beta\} = \{0.46, 0.5\text{Gyr}, 0.03, 0.2, -0.01, 2.4\}$ (navy) and $\{0.46, 1\text{Gyr}, 0.03, 0.2, -0.01, 2.4\}$ (gold), respectively. In both cases, the GSMF and $M_{\bullet}-M_*$ relation are randomly sampled within observational constraints as listed in Tables~\ref{tab:GSMF}-\ref{tab:scaling_relation}. The resulting posterior distributions of the model parameters are shown in Fig.~\ref{Figure.posteriors_delay_fraction}. 
We find that in both scenarios, the parameter $f_3$ remains poorly constrained, whereas the other parameters, $\tau$ and ${\boldsymbol{\Lambda}}_{\text{dN/dt}}$, are well determined. The flat posterior of $f_3$ arises from its strong degeneracy with the delay time and the galaxy merger rate (at $z > 3$). Although the model is sensitive to $f_3$ at $z > 3$ when $\tau$ and ${\boldsymbol{\Lambda}}_{\text{dN/dt}}$ are fixed, the presence of strong degeneracies implies that the model primarily responds to specific combinations of these parameters rather than to $f_3$ individually. 

\begin{figure} 
\centering 
\includegraphics[width=0.45\textwidth]{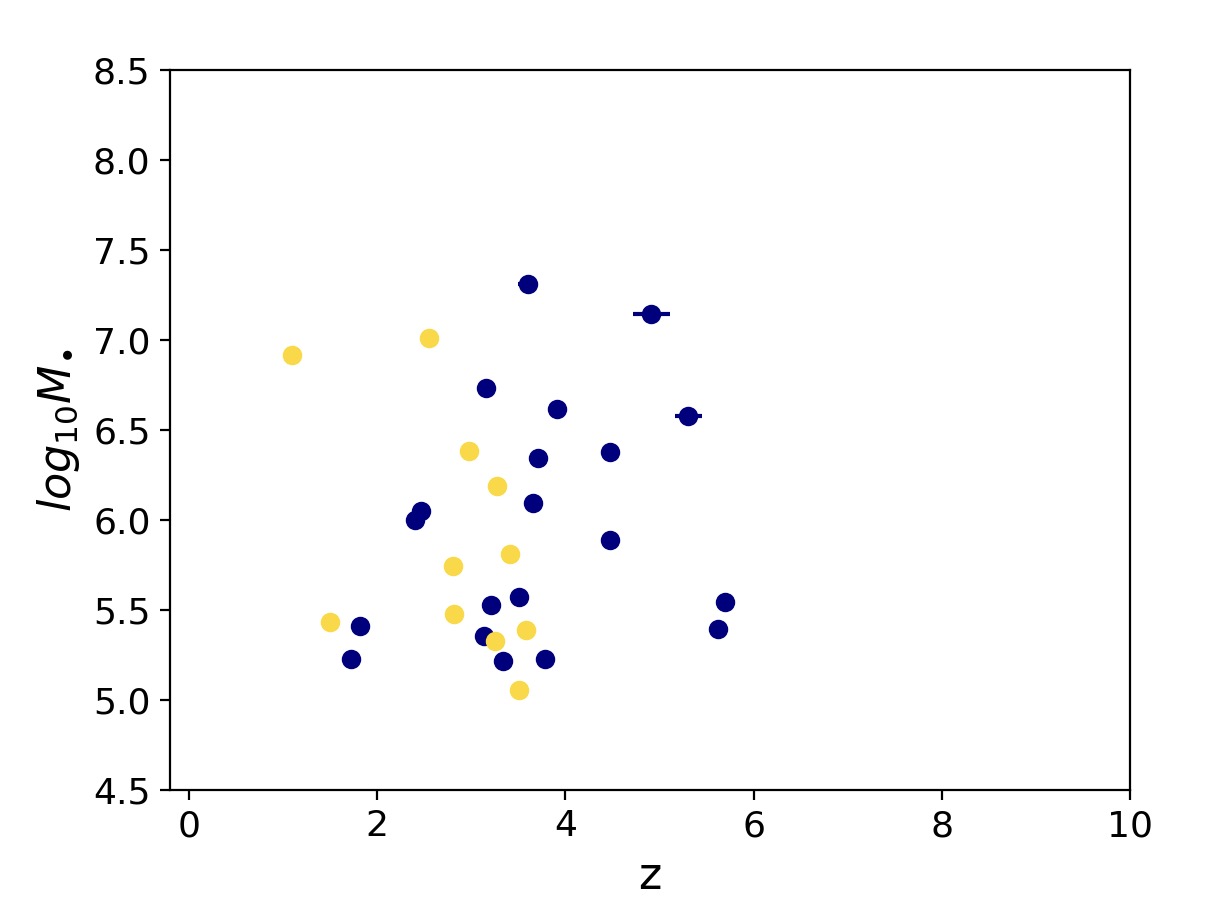}  
\caption{
Mock LISA data realized from merger rate model assuming 
$\{\tau, f_3, \alpha_0, \alpha_1, \beta\} = \{0.5\text{Gyr}, 0.46, 0.2, -0.01, 2.4\}$ (navy) and $\{\tau, f_3, \alpha_0, \alpha_1, \beta\} = \{1\text{Gyr}, 0.46, 0.2, -0.01, 2.4\}$ (gold) respectively, and the GSMF and $M_{\bullet}-M_*$ relationship are set to Tables~\ref{tab:GSMF}-\ref{tab:scaling_relation}.  
} 
\label{Figure.GW_data_delay_fraction} 
\end{figure}

\begin{figure} 
\centering 
\includegraphics[width=0.47\textwidth]{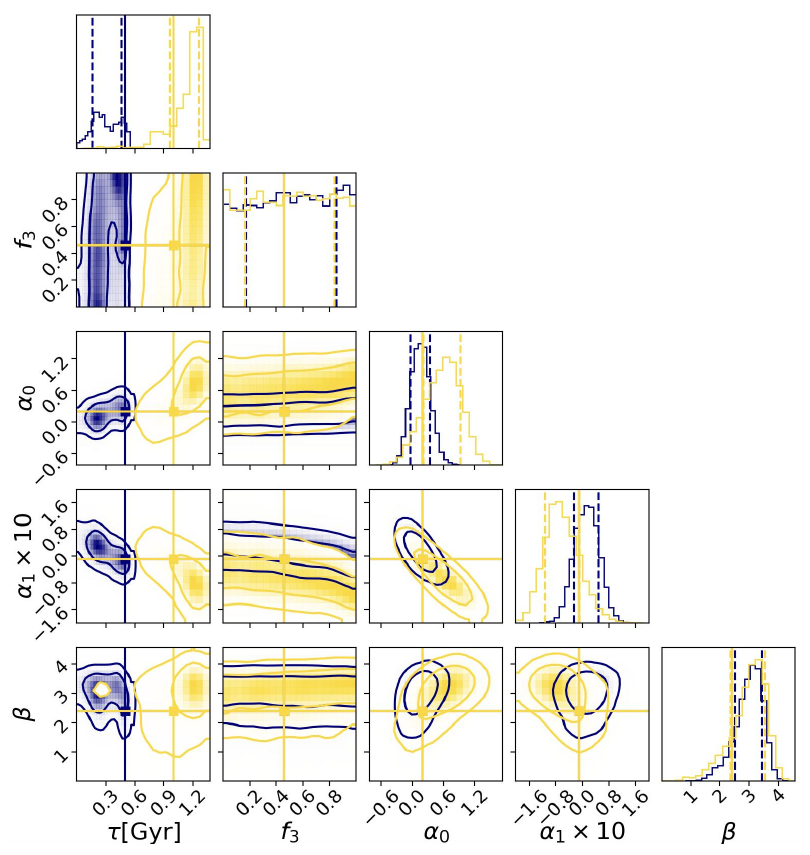}  
\caption{
Posteriors of model parameters ${\bf \Lambda}=\{\tau, f_3, \alpha_0, \alpha_1, \beta\}$, inferred from the mock LISA data shown in Fig.~\ref{Figure.GW_data_delay_fraction}, assuming delay time of $\tau =0.5 \text{Gyr}$ (navy) and $1\text{Gyr}$ (gold). The inner and outer lines show the 0.68 and 0.954 levels. The horizontal and vertical lines represent the corresponding injected parameter values. 
} 
\label{Figure.posteriors_delay_fraction} 
\end{figure}

\section{Conclusion and discussion}
\label{conclusion}
This work  investigates the merger rate of SMBHs and their host galaxies using SGWB detections from PTAs and mock GW data for LISA-like detectors. The findings highlight the critical role of GW detections with LISA-like detectors together with PTAs in exploring the galaxy/SMBH mergers in the hierarchical assembly and the mass growth of SMBHs. 
By incorporating observational constraints from the $M_{\bullet}-M_*$ relation \citep[][]{Kormendy:2013dxa,Pacucci_2024} and GSMFs \citep[][]{Baldry2012, HuertasCompany2016, Santini2012, McLeod2021, Song2016, Stefanon2021}, this study provides a framework for estimating the merger rate of SMBHs and their host galaxies.  

The PTAs and LISA-like detectors provide complementary windows into the SMBH merger process. The current PTAs provide a piece of evidence of a SGWB \citep[][]{NANOGrav_2023, EPTA:2023sfo, EPTA:2023fyk, Reardon_2023, CPTA_2023}, which is most likely to be sourced by SMBH binaries of mass $10^{8}-10^{9}\Msun$ dominated by redshift $z<3$. 
The future LISA-like detector is capable of detecting SMBH binary merger events at higher redshifts ($z=20-30$) and wide mass range of SMBHs ($10^{5}-10^{8}\Msun$). 
The results demonstrate that LISA’s detection of SMBH binary mergers is crucial for reconstructing merger rates (Fig.~\ref{Figure.posteriors_case1_eg1_eg2}-\ref{Figure.recover_galaxy_rate_eg1_eg2}), the delay time of SMBH binary mergers (Fig.~\ref{Figure.posteriors_delay}), and revealing the contribution of mergers to SMBH growth (Fig.~\ref{Figure.SMBH_mass_assemble}).
Moreover, incorporating PTA constraints on the SGWB further refines model parameters, reducing uncertainties.

A key outcome of this study is the ability to compare galaxy merger rate at different redshift bins inferred from GW detections with those obtained from galaxy pair observations \citep[e.g.][]{Casteels2014, Sanjuan2015, Duncan_2019, Conselice_2022, DuanQiao2024} and cosmological simulations \citep[e.g.][]{Vicente2015} (seeing Fig.~\ref{Figure.recover_galaxy_rate_eg1_eg2} for details). GW detections provide independent approaches to estimating galaxy merger rates and thus serve as a valuable cross-check for one another.
Our analysis using mock LISA data indicates that the number of detected events and their joint distribution in mass and redshift are crucial for constraining galaxy and SMBH merger rates. Specifically, GW data with 36 events yield results consistent with galaxy pair observations, whereas data with only 12 events fail to capture the high-redshift behavior. At higher redshift, larger biases emerge due to the limited number of detections and increased uncertainties in the GMSFs and the $M_{\bullet}$–$M_*$ relations. 

SMBHs assemble their mass through mergers and accretion.  
The comparison between these two processes could reveal the mass assembly histories of SMBHs. 
In our analysis (Fig.~\ref{Figure.SMBH_mass_assemble}), the recovered merger mass assembly at high redshift (particularly at $z\ge \ 4$) exhibits significant uncertainties, this is mainly caused by the uncertainties in the SMBH mass function propagated from the $M_{\bullet}-M_*$ relation and GSMFs, as well as the uncertainties in the estimated galaxy merger rate at such redshift bin. 
Future observations providing more robust constraints on the $M_{\bullet}-M_*$ relation and GSMFs at high redshifts will help reduce these uncertainties. 
The delay time between galaxy mergers and subsequent SMBH coalescences suppresses the SMBH merger rate and can significantly affect the contribution of mergers to the growth of SMBHs. 
Our analysis (Fig.~\ref{Figure.SMBH_mass_assemble}) shows that in models with short or vanishing delay times ($\tau\le 0.2\text{Gyr}$), mergers typically dominate SMBH mass assembly at high redshifts ($z > 4$). 
In contrast, for models with delay times longer than $0.5\text{Gyr}$ ($0.8\text{Gyr}$), accretion becomes the primary driver of SMBH mass growth beyond $z \sim 6$ ($4$). 

We further examine the estimation of the SMBH occupation fraction in galaxies at $z > 3$ ($f_3$), parameterized similarly to \citet{Beckmann2023} and \citet{Langen2025}, in conjunction with the delay time and galaxy merger rate parameters (Fig.~\ref{Figure.posteriors_delay_fraction}). While the delay time and merger rate parameters are well constrained, $f_3$ remains poorly determined. Its flat posterior distribution reflects its significant degeneracies with both the delay time and the galaxy merger rate. The delay time parameter is properly constrained (Figs.~\ref{Figure.posteriors_delay} and \ref{Figure.posteriors_delay_fraction}), consistent with our previous findings \citep[][]{Fang:2022cso}, where both power-law and Gaussian delay time distributions could be recovered. In this work, due to computational limitations, we restrict our analysis to a constant delay model. Future studies in this framework should explore more realistic delay time models described by distribution functions to better capture the complexities of SMBH merger dynamics. 

In summary, this study demonstrates the strengths and limitations of combining GW detections with observational constraints on scaling relations and GSMFs to investigate galaxy and SMBH merger rates, estimate the coalescence time of SMBH binaries and its role in the SMBH mass assembly through mergers, and constrain the SMBH occupation fraction in galaxies. 
Future advancements in GW observatories, such as LISA \citep[][]{eLISA:2013xep, LISA_2017}, Taiji \citep[][]{YueliangWu_taiji}, Tianqin \citep[][]{luo2016tianqin} , and PTAs \citep[][]{NANOGrav_2023, EPTA:2023sfo, EPTA:2023fyk, Reardon_2023, CPTA_2023}, alongside improved electromagnetic observations from JWST \citep[e.g.][]{Schneider_2023, matthee2024environmental}, ngEHT \citep[e.g.][]{Orazio_2018, Fang:2021xab, ngEHT_2023} and other facilities will provide deeper insights into the formation and coevolution of SMBHs and their host galaxies. These efforts will ultimately enhance our understanding of the large-scale structure of the universe and the role of SMBH/galaxy mergers in shaping their evolution.

\section*{Acknowledgements}
We acknowledge the use of the HPC Cluster of the National Supercomputing Center in Beijing. This work makes use of the open-sourced python package emcee \citep{Foreman2013emcee}. We thank Youjun Lu and Yunfeng Chen for useful discussions. We also thank the anonymous referee for the constructive suggestions that helped improve the quality of this work. 
YF is supported by the National Natural Science Foundation of China (Grant No. 12405068) and the Start-up Research Fund Project at Ningbo University. 
RGC is supported by the National Natural Science Foundation of China with Grant No. 12235019.

\section*{Data availability}
Data is available upon reasonable request from the authors.

\bibliographystyle{mnras}
\bibliography{reference}
%
\appendix
\section{Settings of merger rate model for generating mock GW data}
\label{sec.app_A}
\begin{figure*}
\centering 
\includegraphics[width=0.75\textwidth]{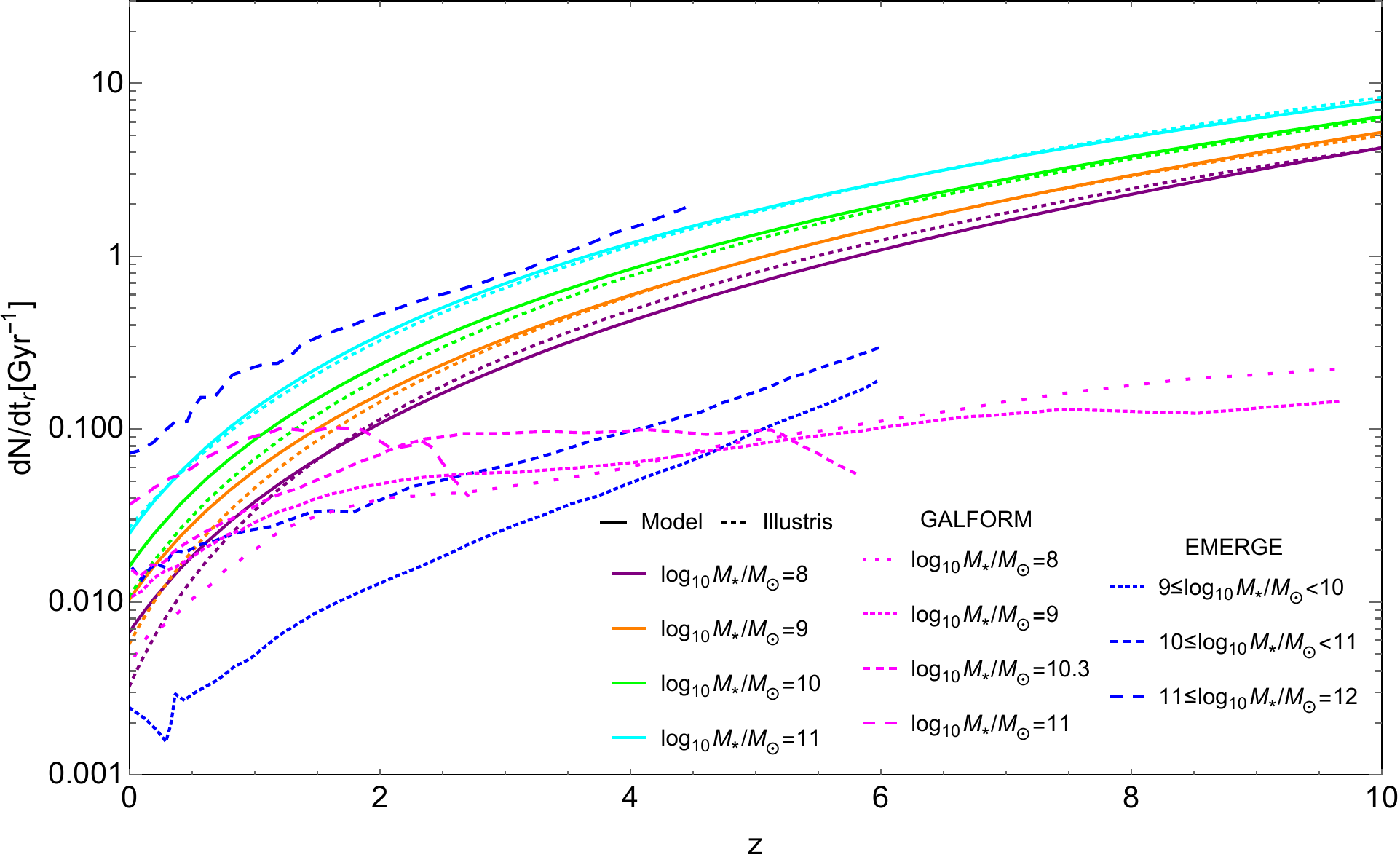}  
\caption{
Galaxy merger rates per galaxy assumed in our model for generating the mock LISA GW data (solid lines), compared with predictions from the Illustris simulations (colored dashed lines, \citep[][]{Vicente2015}), GALFORM (magenta dashed lines, \citep[][]{Filip_2022_galaxy}), and EMERGE (blue dashed lines, \citep[][]{OLeary_2021}).  
} 
\label{Figure.dN_dtr} 
\end{figure*}

In the appendix, we detail the settings of the merger rate model used to generate the mock GW data discussed in Sections \ref{infer_merger_rate}, \ref{sec_delay_time}, and \ref{sec_occupation}. Specifically, Fig.~\ref{Figure.dN_dtr} shows the galaxy merger rate per galaxy assumed in our model, alongside a comparison with predictions from the simulations in Illustris \citep[][]{Vicente2015}, GALFORM \citep[][]{Filip_2022_galaxy}, and EMERGE \citep[][]{OLeary_2021}. 
The GSMFs and $M_{\bullet}-M_*$ relation at different redshift bins are sampled from observational constraints, as listed in Tables.~\ref{tab:GSMF} and \ref{tab:scaling_relation}.  

%

\begin{table*}
\centering
\caption{
Parameters $\[ \mathcal{M}_{\star}, \log_{10} \phi_{1}, a_{1}, \log_{10}\phi_{2}, a_{2}\]$ of the galaxy stellar mass function (equation \ref{GSMF_params}) across different redshift bins, sampled from observational constraints. These values are used to generate the mock LISA data discussed in Sections~\ref{infer_merger_rate}, \ref{sec_delay_time}, and \ref{sec_occupation}. }
\label{tab:GSMF}
\begin{tabular}{ c c c c}
\hline
Redshift & GW example & $\[ \mathcal{M}_{\star}, \log_{10} \phi_{1}, a_{1}, \log_{10}\phi_{2}, a_{2}\]$ & bins of redshift \\
\hline
\multirow{4}{*}{$0 < z \leq 3.5$} &  \multirow{6}{*}{Sections \ref{infer_merger_rate}, \ref{sec_delay_time}  \ref{sec_occupation}}  & [[10.66, -2.40, -0.35, -3.10, -1.47 ],  & $[ 0 < z \leq 0.2$, \\
                                        &               &  [10.86, -2.65, -0.82, -3.35, -1.60],  & $0.2 < z \leq 0.8$, \\
                                        &               &  [11.03, -2.88, -1.25, -inf, 0],           & $0.8 < z \leq 1.1$, \\
\citep[][]{Baldry2012, Santini2012}  &            &  [11.32, -3.41, -1.42, -inf, 0], &  $1.1 < z \leq 2.0$, \\
\citep[][]{HuertasCompany2016}  &             &  [10.77, -3.18, -0.68, -3.84, -1.73],  & $2.0 < z \leq 3.0 $, \\
\citep[][]{McLeod2021}    &               &  [10.84, -3.94, -1.79, -4.3, 0]] & $3.0 < z \leq 3.5$] \\
                                       &               &                &  \\
\hline
\multirow{20}{*}{$3.5 < z \leq 10.5$} & \multirow{7}{*}{Section \ref{infer_merger_rate}, GW (eg1)} & [[10.77,  -3.78, -1.59,   -inf,  0 ], &  \\
                                         &                                        &  [11.08,  -4.20, -1.61,   -inf,  0 ],   & \\
                                         &                                        & [10.45,  -4.45, -1.82,    -inf,  0 ],   & \\
                                        &                                        & [10.17,  -4.27, -1.83,    -inf,  0 ],    & \\
                                        &                                        & [10.82,  -4.74, -1.38,    -inf,  0 ],     & \\
                                         &                                        & [ 9.50,   -5.18, -2.00,    -inf,  0 ],    & \\
                                         &                                        & [ 9.50,   -6.52, -2.00,    -inf,  0 ]]    &  \\
& & & \\                           
                    & \multirow{7}{*}{Section \ref{sec:5.1}, GW (eg2)} &[[10.91,  -3.87, -1.55,  -inf,  0 ], & [$3.5 < z \leq 4.5$, \\
                                         &                                        & [10.66,  -4.07, -1.65,     -inf,  0 ],  & $4.5 < z \leq 5.5$, \\
                                         &                                        & [ 9.90,  -3.94,  -1.80,     -inf,  0 ],  &  $5.5 < z \leq 6.5$, \\
\citep[][]{Song2016}          &                                        & [10.47,  -5.27, -2.03,     -inf,  0 ],  & $6.5 < z \leq 7.5$, \\
\citep[][]{Stefanon2021}    &                                        & [ 9.42,  -4.05,  -1.83,     -inf,  0 ],   & $7.5 < z \leq 8.5$, \\
                                         &                                        & [ 9.50,  -5.05,  -2.00,     -inf,  0 ],    & $8.5 < z \leq 9.5$, \\
                                         &                                        & [ 9.50,  -6.26,   -2.00,    -inf,  0 ]]   & $9.5 < z \leq 10.5$] \\
& & & \\                                        
                   & \multirow{7}{*}{Sections \ref{sec_delay_time},  \ref{sec_occupation}} &[[10.66,  -3.72,  -1.61,  -inf,  0 ], &  \\
                                         &                                        & [11.07,  -4.23,  -1.65,     -inf,  0 ],  & \\
                                         &                                        & [10.58,  -4.50, -1.80,      -inf,  0 ],  &  \\
                                         &                                        & [10.29,  -4.24, -1.67,      -inf,  0 ],   & \\
                                         &                                        & [ 8.87,  -2.85,  -0.77,      -inf,  0 ],    & \\
                                         &                                        & [ 9.50,  -5.21,  -2.00,      -inf,  0 ],  &  \\
                                         &                                        & [ 9.50   -5.79,  -2.00,      -inf,  0 ]]  &  \\
\hline
\end{tabular}
\end{table*}
\begin{table*}
\centering
\caption{
Parameters $\[ a, b, \epsilon \]$ of the $M_{\bullet}-M_*$ relationship (equation (\ref{scaling_relation})), sampled from observational constraints, used in generating the mock LISA data discussed in Sections \ref{infer_merger_rate}, \ref{sec_delay_time}, and \ref{sec_occupation}.}
\label{tab:scaling_relation}
\begin{tabular}{ c c c}
\hline
Redshift & GW example & $\[ a, b, \epsilon \]$ \\
\hline
\multirow{4}{*}{$0 < z \leq 4$}        &   Section \ref{infer_merger_rate}, GW (eg1)            &  [-5.36, 1.28, 0.28]       \\
& & \\
                                                   &  Section \ref{infer_merger_rate}, GW (eg2)              &  [-6.10, 1.35, 0.28]        \\
\citep[][]{Kormendy:2013dxa}                 & & \\
                                                   &   Sections \ref{sec_delay_time},  \ref{sec_occupation}             &  [-4.25, 1.18, 0.28]        \\
\hline
\multirow{4}{*}{$4 < z \leq 10.5$}          &   Section \ref{infer_merger_rate}, GW (eg1)             &   [-2.53, 1.02, 0.69 ]        \\
& & \\
                                                  &  Section \ref{infer_merger_rate}, GW (eg2)              &   [-3.04, 0.96, 0.69 ]        \\
\citep[][]{Pacucci_2024}             & & \\
                                                  &   Sections \ref{sec_delay_time},  \ref{sec_occupation}             &   [-1.93, 1.05, 0.69 ]         \\
\hline
\end{tabular}
\end{table*}

\section{Galaxy merger rate estimated assuming local scaling relationship}
\label{sec.app_B}
\begin{figure*}
\centering 
\includegraphics[width=0.8\textwidth]{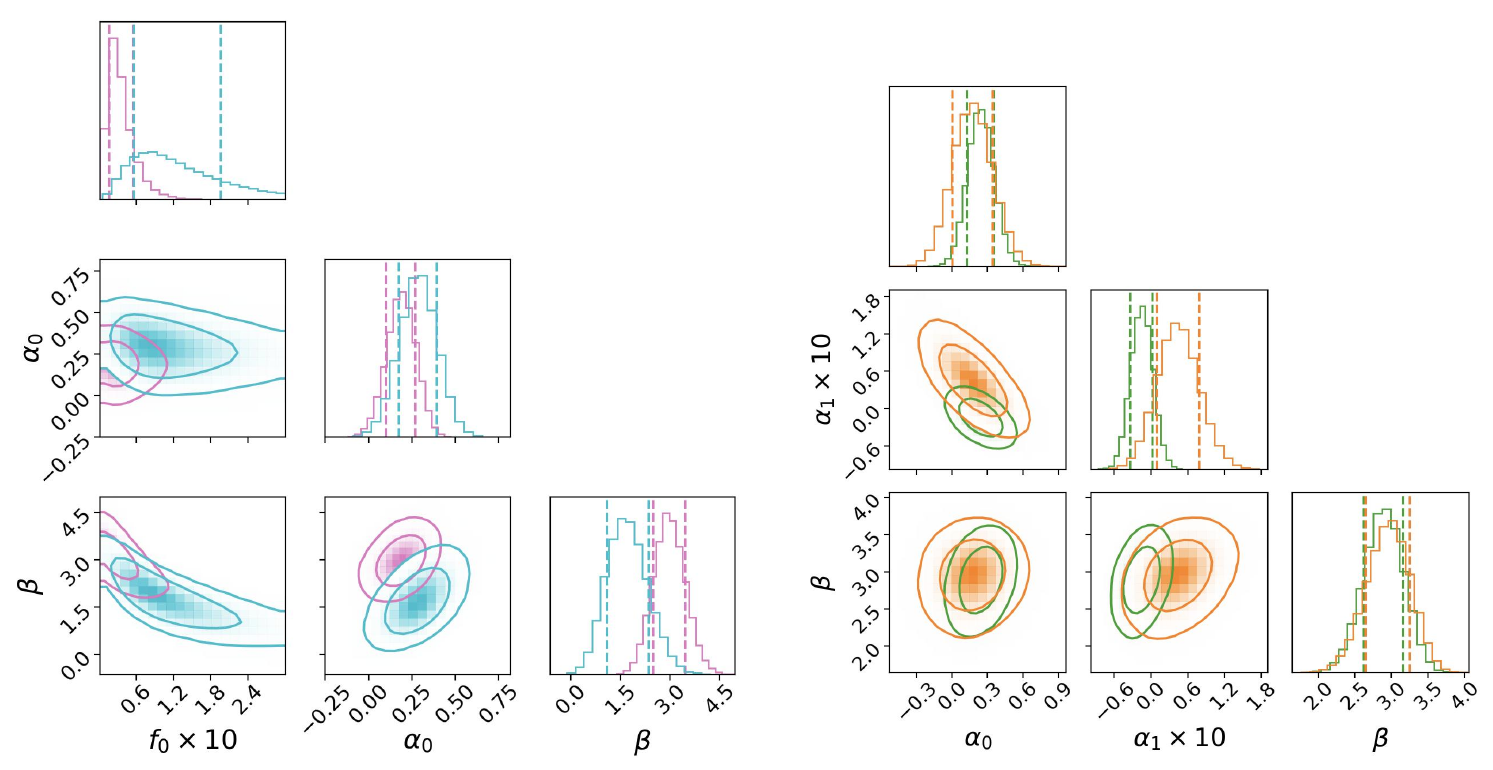}  
\caption{
Left: posteriors of the model parameters in Case 1 model ($dN/dt=$) estimated from the first  (pink) and second (cyan) realization of mock LISA data shown in Fig.~\ref{Figure.LISA_data}, together with PTA detections of SGWB strain (Fig.~\ref{Figure.hc}). 
Right: posteriors of the model parameters in Case 2 model ($dN/dt=$) estimated from the first  (green) and second (orange) realization of mock LISA data shown in Fig.~\ref{Figure.LISA_data}, together with PTA detections of SGWB strain (Fig.~\ref{Figure.hc}). 
} 
\label{Figure.posteriors_local_relation} 
\end{figure*}

\begin{figure*} 
\centering 
\includegraphics[width=0.95\textwidth]{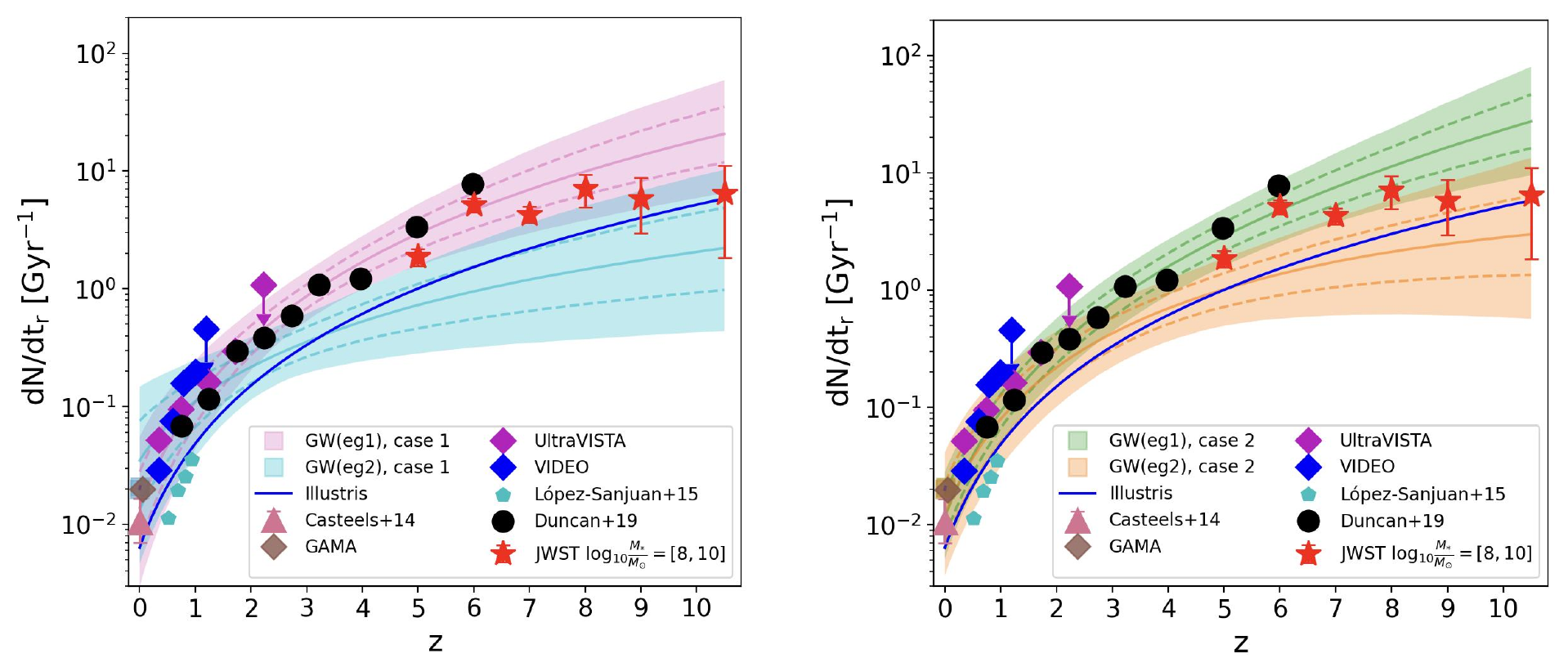}  
\caption{
Similar to Fig.~\ref{Figure.recover_galaxy_rate_eg1_eg2} but for Galaxy merger rate ${dN\over dt_{\rm r}}(z)$ estimated assuming local $M_{\bullet}-M*$ relationship \citep{Kormendy:2013dxa} across the redshifts. The line styles in the left and right hand panels are the same to Fig.~\ref{Figure.recover_galaxy_rate_eg1_eg2}. The one sigma (bounded by dashed lines) and two sigma (shaded region) confidence region is plotted according the posteriors results shown in Fig.~\ref{Figure.posteriors_local_relation}.
} 
\label{Figure.recover_galaxy_merger_rate_local_relation} 
\end{figure*}

In the appendix, we present the posterior distributions of model parameters in the Case 1 and Case 2 models in Fig.~\ref{Figure.posteriors_local_relation}, and the reconstructed galaxy merger rate per galaxy from those posteriors in Fig.~\ref{Figure.recover_galaxy_merger_rate_local_relation}, by using the local $M_{\bullet}-M_*$ relationship \citep[][]{Kormendy:2013dxa} and the GW datasets discussed in Section \ref{infer_merger_rate}. 
%
\label{lastpage}
\end{document}